# Revealing the impact of polystyrene-functionalization of Au octahedral nanocrystals of different sizes on formation and structure of mesocrystals

*Dmitry Lapkin[+], Shweta Singh, Felizitas Kirner, Sebastian Sturm, Dameli Assalauova, Alexandr Ignatenko, Thomas Wiek, Thomas Gemming, Axel Lubk, Knut Müller-Caspary, Azat Khadiev, Dmitri Novikov, Elena V. Sturm\*, Ivan A. Vartanyants\**

Dmitry Lapkin, Shweta Singh, Dameli Assalauova, Alexandr Ignatenko, Azat Khadiev, Dmitri Novikov, Ivan A. Vartanyants
Deutsches Elektronen-Synchrotron DESY, Notkestr. 85, D-22607 Hamburg, Germany
E-mail: ivan.vartaniants@desy.de

Felizitas Kirner, Elena V. Sturm
Department of Earth and Environmental Sciences, Section of Crystallography,
Ludwig Maximilian University of Munich (LMU), Theresienstr. 41C, 80333 Munich, Germany
E-mail: sturm.elena@lmu.de

Sebastian Sturm, Knut Müller-Caspary
Department of Chemistry and Centre for NanoScience, Ludwig-Maximilians-Universität München (LMU), Butenandtstr. 11, 81377 Munich, Germany

Thomas Wiek, Thomas Gemming, Axel Lubk
Leibniz Institute for Solid State and Materials Research, Helmholtzstraße 20, 01069 Dresden, Germany

[+]Dmitry Lapkin
*Current address: Institute of Applied Physics, University of Tübingen, Auf der Morgenstelle 10, D-72076 Tübingen, Germany*

**June 26, 2023**

**Keywords:** Gold octahedral nanoparticles; mesocrystals; angular X-ray cross-correlation analysis

**Abstract**

The self-assembly of anisotropic nanocrystals (stabilized by organic capping molecules) with pre-selected composition, size, and shape allows for the creation of nanostructured materials with unique structures and features. For such a material, the shape and packing of the individual nanoparticles play an important role. This work presents a synthesis procedure for ω-thiol-terminated polystyrene (PS-SH) functionalized gold nanooctahedra of variable size (edge length 37, 46, 58, and 72 nm). The impact of polymer chain length (Mw: 11k, 22k, 43k, and 66k g·mol$^{-1}$) on the growth of colloidal crystals (*e.g.* mesocrystals) and their resulting crystal structure is investigated. Small-angle X-ray scattering (SAXS) and scanning transmission electron microscopy (STEM) methods provide a detailed structural examination of the self-assembled faceted mesocrystals based on octahedral gold nanoparticles of different size and surface functionalization. Three-dimensional angular X-ray cross-correlation analysis (AXCCA) enables high-precision determination of the superlattice structure and relative orientation of nanoparticles in mesocrystals. This approach allows us to perform non-destructive characterization of mesocrystalline materials and reveals their structure with resolution down to the nanometer scale.

*1. Introduction*

In nanotechnology and materials science, the bottom-up production of nanoparticles with regulated structure and elaborate functionality is constantly a conceptual challenge.[1, 2] The self-assembly of anisotropic nanocrystals (stabilized by organic capping molecules) with pre-selected composition, size, and shape allows for the creation of nanostructured materials with unique structures and features. Colloidal crystals, especially with the defined crystallographic orientation of the nanoparticles (*e.g.* mesocrystals), can demonstrate newly emerging and novel collective properties that cannot occur in any other structure of the same size range.[3] For such a material, the shape and packing of the individual nanoparticles plays an important role.[4-6] The arrangement and orientation of nanoparticles in superlattices depend decisively on their shape.[7-10] Theory predicts the assembly of hard octahedral particles (as non-space filling polyhedra) in a wide variety of fascinating crystalline and liquid-crystalline phases with different orientational and translational orders.[11-13] Furthermore, the rounded corners and curved facets of octahedron-like superballs could play an important role in stabilization of specific assembled phases.[14] In the case of regular hard octahedra, simulation finds the densest packing with a density of 18/19 to be the *Minkowski* lattice.[8, 11-13] With increasing truncation, the packing fraction gradually decreases and the packing switches to a *bcc* or *bct* structure.[11, 15] In experimental work, monoclinic and simple hexagonal superlattices are additionally found as possible packing under different assembly conditions.[8] This emphasizes the high adaptability of the superlattice symmetry on the specific crystallization conditions and functionalization (e.g., type of organic capping molecules) of octahedral particles. The functionalization plays an essential role in the self-assembly of nanoparticles from dispersion due to the interaction of organic ligands between each other and solvent molecules.[16] Nevertheless, most simulation studies do not account for the ligand shell around the inorganic core and are usually performed only with respect to the hard nanoparticle core, focusing on the study of specific features or specific implementations of mesocrystals.[17-20] Some computational simulations consider the influence of an elastic organic matrix/ligand (i.e., DNA molecules) on the packing arrangement of the building superstructures, yet, only with limited success.[21] Experimentally, it was shown that the self-assembly of Ag nanocubes in 1D and 2D arrays can be tuned by adjusting the surfactant shell.[22, 23] The ligand conformation may be subjected to changes during the assembly process as environmental conditions may change and thus can significantly differ from both dispersed and assembled states, making prognoses about an assembly process and the outcome structure quite complex.

In this work, we studied gold nanoparticles with octahedral shape stabilized by polymer macromolecules with a specific chain length, which not only defines the interparticle distance in assembly, but also affects the effective shape of the nanoparticles, particle-particle, and particle-solvent interactions. This provides an excellent playground system to create ordered plasmonic self-assemblies with potentially tunable structures and properties. We herein present a synthesis procedure for ω-thiol-terminated polystyrene (PS-SH) functionalized gold nanooctahedra of variable size (edge length 37, 46, 58, and 72 nm) and investigate the impact of the polymer chain length (Mw: 11k, 22k, 43k, and 66k g·mol$^{-1}$) on the growth of colloidal crystals (*e.g.* mesocrystals) and their resulting crystal structure. Small-angle X-ray scattering (SAXS) and advanced X-ray and electron imaging methods provide a complete structural examination of the self-assembled superstructures based on nanoparticles of various compositions and forms.[20, 24-29] Angular X-ray cross-correlation analysis (AXCCA), which was recently developed, enables high-precision determination of colloidal crystal structure and relative orientation of nanoparticles in mesocrystals.[30-34] Therefore, in this publication, we combine these two approaches and measure a large part of three-dimensional (3D) reciprocal space in SAXS geometry, including Bragg peaks from the superlattice structure, for individual grains. The AXCCA of the measured 3D scattered intensity distributions allows revealing the structure of the mesocrystalline grains with a resolution down to the nanometer scale.

## 2. Results and discussion

### *2.1. Synthesis of PS-SH functionalized gold octahedral nanoparticles and growth of colloidal crystals*

Octahedral nanocrystals (37 – 72 nm edge length) are synthesized in aqueous dispersion according to the literature[35]. These nanoparticles are successfully functionalized by ω-thiol-terminated polystyrene (PS-SH) of different molecular weights (11k g·mol$^{-1}$ - 66k g·mol$^{-1}$) within the process of a phase transfer to toluene (detailed synthesis parameters can be found in the Supporting Information).[36] UV/Vis spectra before and after functionalization (**Figure S3**, Supporting Information) are recorded to ensure the retention of shape and dispersed state after the phase transfer, no peak broadening is found indicating such processes. Transmission electron microscopy (TEM) investigation of these polystyrene-functionalized particles (**Figure 1a**) further confirms the shape retention and point to an influence of the polymer weight on the packing of the dried nanoparticles. TEM images also indicate that for such small di- or trimers, longer polymer chains (e.g., higher molecular weight) lead to larger distances between the nanocrystal cores and a more random orientation of the nanocrystals towards each other. However, the size of the nanoparticle core also seems to influence the orientation. The shape of the individual gold nanocrystals was verified by high-angle annular dark-field imaging (HAADF) in a scanning transmission electron microscopy (STEM) tomography (**Figure 1 e, Figure S2,** Supporting Information). The degree of truncation along the ⟨100⟩ directions has been estimated to be approximately α = 0.14±0.02, by comparing the deviation of tip-to-tip distances to the ideal octahedron ($\alpha = 1 - d_{t-t}^{real}/d_{t-t}^{ideal}$, where $d_{t-t}^{real}$ and $d_{t-t}^{ideal}$ are the tip-to-tip distance along the ⟨100⟩ directions in a real and ideal octahedra, respectively). The degree of truncation along the ⟨110⟩ directions is below the resolution limit of tomographic reconstruction. These values are also affected by the tomographic reconstruction resolution, namely because of artefacts (*e.g.* due to a "missing wedge") but should give a good estimate of the level of the edge truncation.

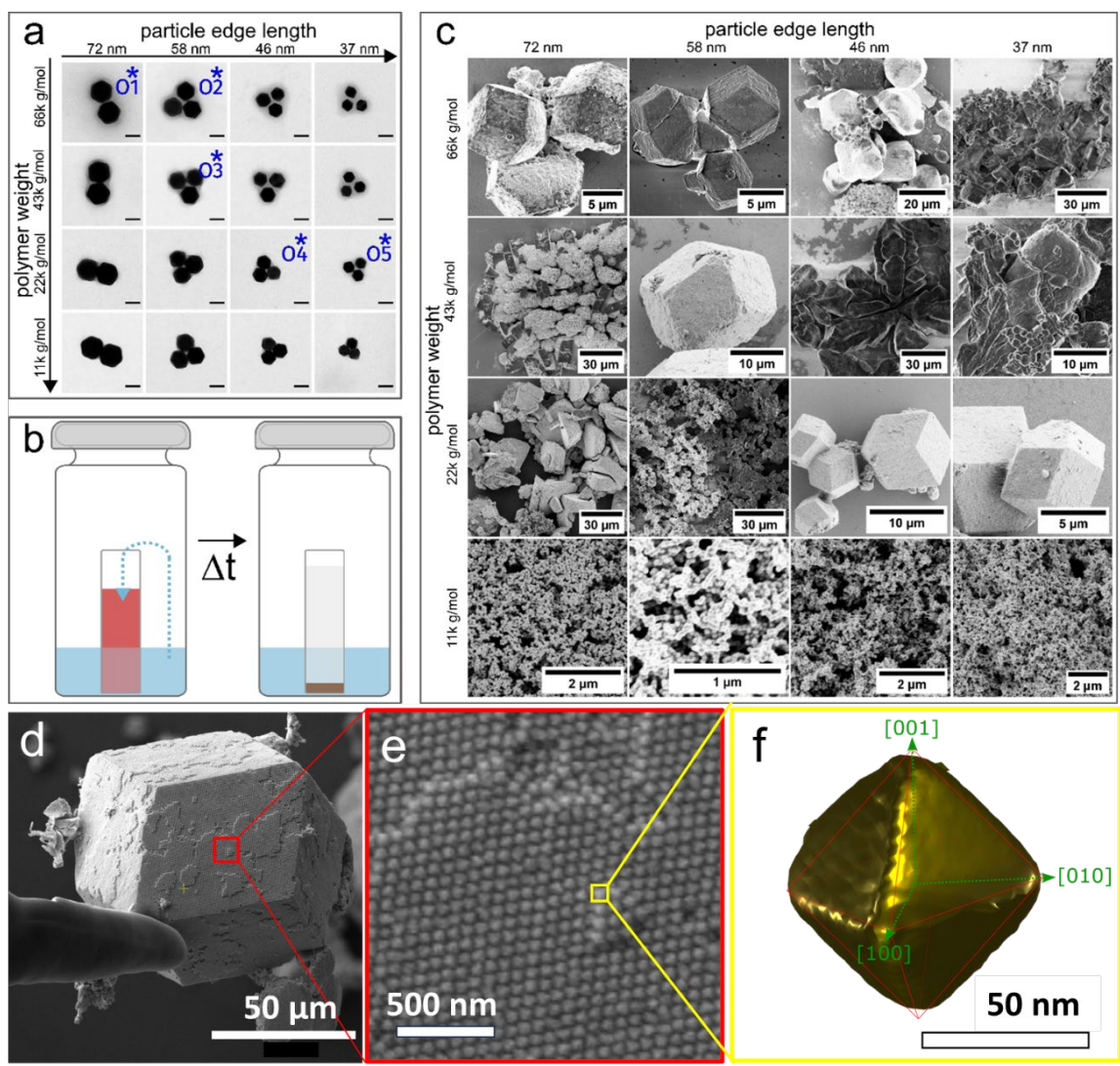

**Figure 1.** The synthesis of mesocrystals includes the synthesis and functionalization of nanoparticles followed by their assembly. a) Several combinations of core nanooctahedra sizes and ligand polystyrene (PS-SH) chain weights are synthesized. TEM images prove that the octahedral shape of the Au cores is retained during the functionalization process and an influence of the polymer weight on the assembly structure can be observed. The mesocrystalline samples assembled from the nanoparticles with the nanooctahedra sizes and polymer molar weights marked by a blue asterisk are investigated by X-ray diffraction. The scale bar is 50 nm. b) Nanoparticle self-assembly using a gas phase diffusion approach. The sketched setup depicts a closed vial containing the antisolvent and a smaller one containing the nanoparticle dispersion. Over time, the antisolvent leads to the destabilization of NPs, self-assembly, and discoloration

of the dispersion. c) Self-assembled superstructures as collected from the bottom of the vial, investigated by SEM. Superstructure size and morphology vary depending on the polymer/core size combination. Facetted colloidal crystals are found for polymer weights of 22k g·mol$^{-1}$ and above. d) Exemplary SEM image of a mesocrystal of rhombic dodecahedral shape assembled from PS-SH capped octahedral nanoparticles (72 nm edge length, 66k g·mol-1 PS-SH). e) The zoomed image shows the surface structure of the mesocrystal and packing of its nanoparticles. f) HAADF-STEM tomographic reconstruction of an exemplary single octahedral gold nanoparticle (for more details see Figure S2, Supporting Information).

Dynamic light scattering (DLS) measurements are conducted to gain deeper insight into the influence of the polystyrene ligand on the apparent size of the particles in the organic solution (**Figure S4**, Supporting Information). It is found that a longer polymer results in a higher hydrodynamic radius of the nanoparticle meaning that the polymer chains extend further into the dispersing agent, sterically stabilizing the nanoparticles. The comparatively narrow polydispersity index (PDI) from these measurements confirms the conclusion from UV/Vis spectra that the particles are fully dispersed in solution and no aggregation is occurring. All combinations of particle sizes and polymer weights are fully stabilized.

The self-assembly of the particles is induced using a gas-phase diffusion technique with an ethanol mixture as antisolvent (**Figure 1b**).[24] The antisolvent diffuses over the gas phase into the nanoparticle-toluene dispersion and gradually decreases the solvent quality for the stabilizing polystyrene, leading to destabilization, and self-assembly of the particles. The self-assembly is fastest for short polymers and large nanocrystals and slowest for long polymers with small nanocrystals, suggesting a direct influence of the polymer length on the stability of the nanoparticles. Depending on the nanoparticle core size and polymer weight, 3D colloidal crystals or aggregate networks are obtained (**Figure 1c**). For low polymer weight (11k g·mol$^{-1}$), the destabilization seems to be too fast, preventing an ordered assembly of nanoparticles and merely leading to aggregation. For polymer weights of 22k g·mol$^{-1}$ and above, 3D superstructures are built. In general, we observe the following tendency: longer polymers and smaller nanoparticles produce larger superstructures. This emphasizes the crucial influence of the nanoparticle stability on the self-assembly process and the necessity of the nanoparticles to be able to attach and detach to a building superstructure to build nicely defined superstructures. Three-dimensional superstructures with well-defined faceting (rhombic dodecahedral shape) and suitable size are obtained for 5 polymer-nanocrystal combinations (**Figure 1d, Figure S5**, Supporting Information): a) O1 sample (72 nm edge length, 66k g·mol$^{-1}$ PS-SH). b) O2 sample (58 nm edge length, 66k g·mol$^{-1}$ PS-SH). c) O3 sample (58 nm edge length, 43k g·mol$^{-1}$ PS-SH). d) O4 sample (46 nm edge length, 22k g·mol$^{-1}$ PS-SH). e) O5 sample (37 nm edge length, 22k g·mol$^{-1}$ PS-SH). These samples are chosen for further X-ray structure analysis, as it was possible to pick samples with a diameter of approximately 5 – 10 µm as necessary for the X-ray diffraction (XRD) experiment. Thereby, modification or destruction of the sample by e.g. FIB preparation procedures are circumvented. The selected facetted colloidal crystals were mounted onto tungsten needles and glued with SEMGLU glue (Kleindiek Nanotechnik GmbH) using a Zeiss 1540XB CrossBeam focused-ion-beam scanning-electron-microscope (FIB SEM) (see **Figure 1d**, and **Figure S5,** Supporting Information).

*2.2. Structural characterization of faceted mesocrystals*

The XRD experiment (see **Figure 2** for the schematic layout of the X-ray experimental setup and Methods section for further details) was performed on the five selected faceted mesocrystals (**Figure 1** and **Figure S5,** Supporting Information). The 3D reciprocal space maps of two samples composed of the biggest (O1) and smallest (O5) Au nanooctahedra are presented in **Figure 3a–b** (reciprocal space maps for all samples are given in **Figure S6**, Supporting Information). These 3D maps were obtained from the individual two-dimensional (2D) diffraction patterns of the analyzed samples measured at different angles by interpolating them onto a 3D orthogonal grid with a voxel size of 0.0015 nm$^{-1}$. The average radial profiles of the measured scattered intensity distributions and average radial profiles of the intensity between the Bragg peaks for all measured samples are shown in **Figure 3c**. As one can observe from these profiles, with the decreasing size of the nanooctahedra in the samples O1 – O5, the Bragg peaks shift towards the higher *q*-values indicating the shrinkage of the superlattice unit cell of the mesocrystals. This is valid in all cases except for the samples O2 and O3 composed of nanooctahedra of the same size. A similar trend is observed for intensities taken between the Bragg peaks. These curves mostly resemble the form-factor of individual Au octahedra, though affected by the structure factor at low *q*-values. As it is well seen from **Figure 3c**, the form factor oscillations have a smaller period for the samples composed of bigger nanooctahedra.

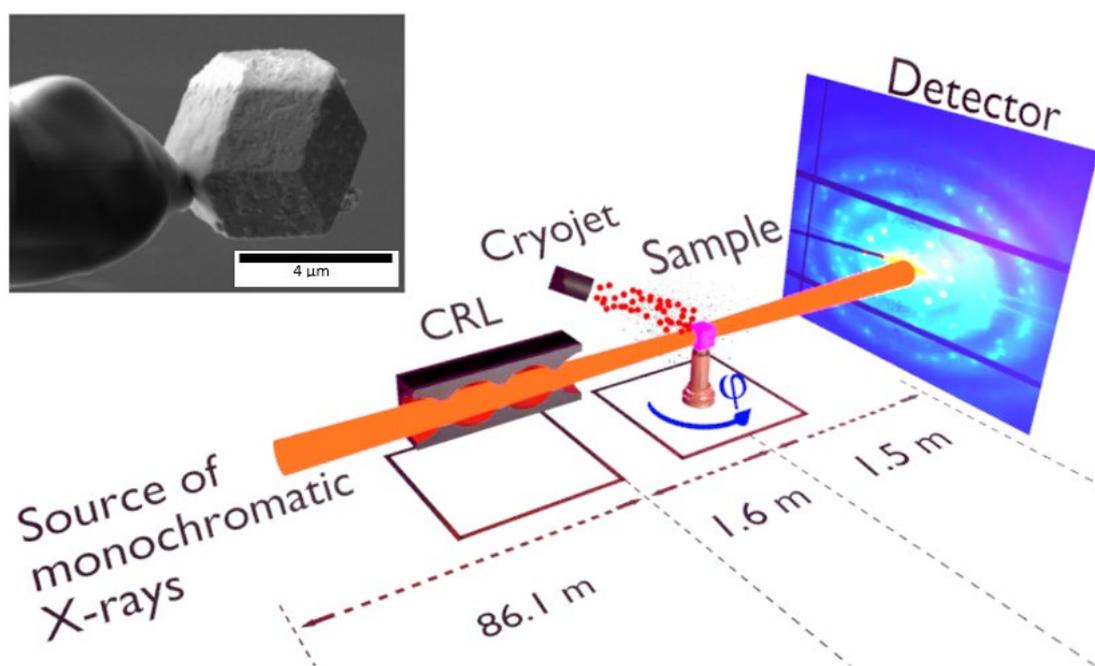

**Figure 2.** The X-ray diffraction experimental setup. Compound Refractive Lenses (CRLs) are used to focus a monochromatic X-ray beam at the sample position. A liquid nitrogen cryojet was used to cool the sample and prevent the radiation damage. An X-Spectrum Lambda 750K detector was positioned downstream at a distance of 1.5 m from the sample to measure the far-field diffraction patterns. The sample was rotated around the vertical axis, and scattering patterns were measured at different angles in the range of $\varphi=0 – 180°$ with the step of $\Delta\varphi=0.5°$. In the inset, an exemplary SEM image of a single mesocrystalline grain fixed on a tungsten tip as measured in the experiment is shown.

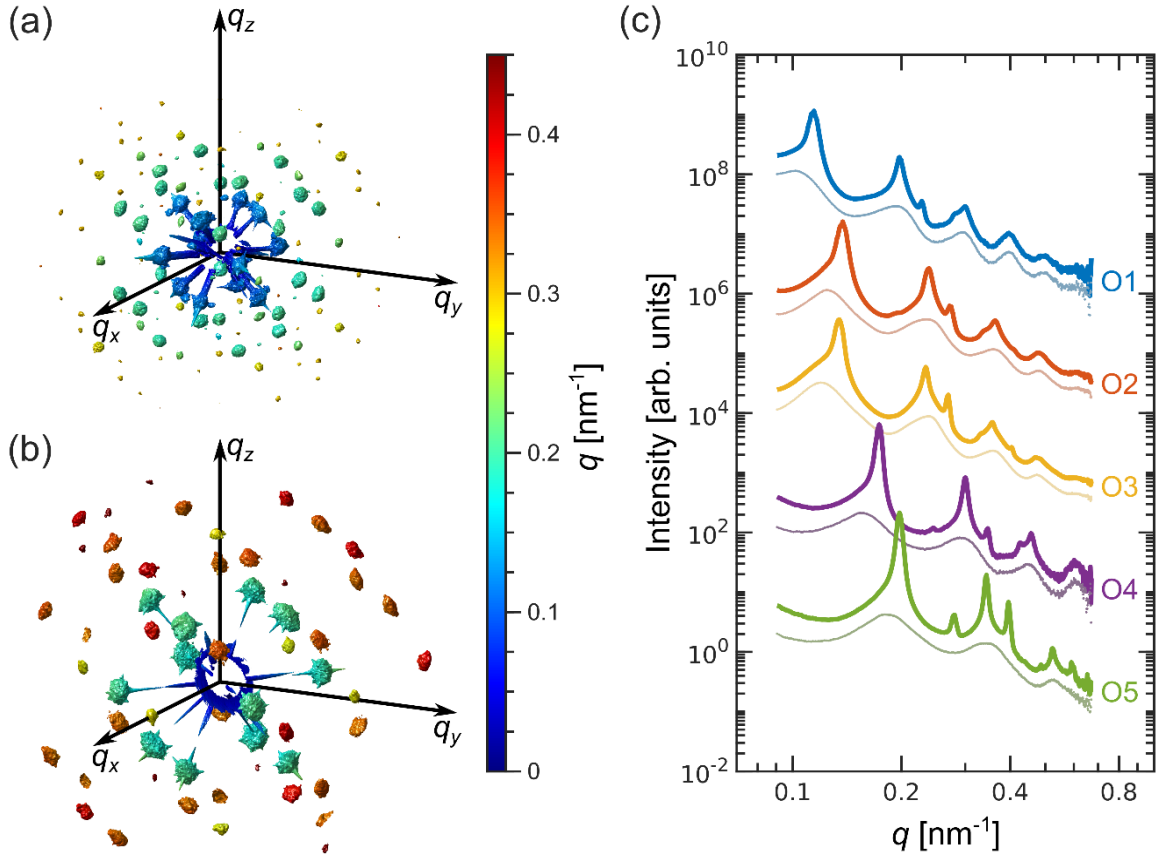

**Figure 3.** (a, b) Examples of the measured scattered intensity distributions in 3D reciprocal space for O1 (a) and O5 (b) mesocrystalline samples. The isosurfaces are rendered for the same arbitrary isovalue. The colors indicate the distance from the origin of the reciprocal space ($q$-values). (c) Average radial profiles of the measured scattered intensity distributions (thick lines) and average radial profiles between the Bragg peaks (thin lines) for all measured samples shown in log-log scale. The resolution is 0.005 nm$^{-1}$. The profiles are shifted vertically for clarity.

To extract the unit cell parameters, we applied AXCCA to the 3D intensity distribution measured for all samples (see Methods section for a description of the AXCCA). Examples of the obtained cross-correlation functions (CCFs) for the samples O1 and O5 are shown in **Figure 4** (the results of AXCCA for all samples are shown in **Figure S7,** Supporting Information). The CCFs $C(q_1,q_2,\Delta)$ were calculated as follows. The value of the first momentum transfer $q_1$ was fixed to the position of the first Bragg peak (see Table 1 for the values) and the value of the second momentum transfer $q_2$ was varied in the range of $q_2 = 0.07 – 0.50$ nm$^{-1}$ with a step size of 0.005 nm$^{-1}$. The CCFs represented in **Figure 4** are shown as functions of $q_2$ and $\Delta$. In this figure, bright yellow spots represent the correlation peaks occurring between the first order Bragg peaks and all others. The CCFs also contain stripe-like features at low $q$-values corresponding to the correlations with the crystal truncation rods produced by {110} facets of rhombic dodecahedral mesocrystalline grains. We note that we did not fully cover reciprocal space for $q > 0.42$ nm$^{-1}$ due to the experimental geometry which explains the absence of some correlation peaks in the CCFs calculated for these $q$-values.

We assumed that the real space structure for all measured samples is body-centered cubic (*bcc*). We performed modeling of the correlation peak positions corresponding to this structure and optimized the unit cell parameter to fit the experimental peak positions (see for details Ref.[37] and section S3.4, Supporting Information). The correlation peak positions for the optimized structures are indicated in **Figure 4** by red circles. The corresponding unit cell parameters for all samples are summarized in **Table 1**.

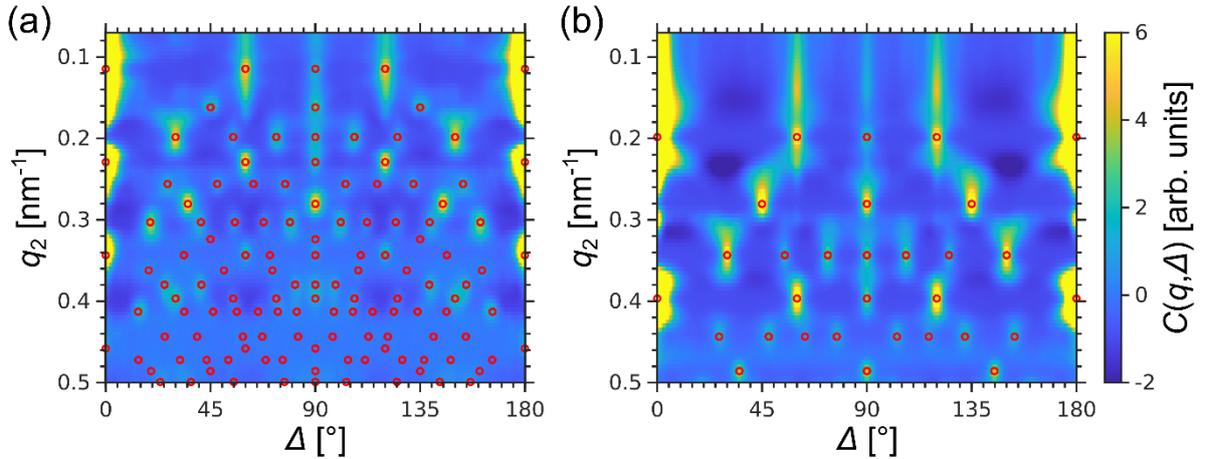

**Figure 4.** 2D maps of cross-correlation functions $C(q_1, q_2, \Delta)$ calculated for the samples O1 (a) and O5 (b). Red circles represent the peak positions for the optimized bcc structure. The cross-correlation functions are calculated for the $q_1$ value corresponding to the first Bragg peak position (see Table 1 for the values) and $q_2$ varying in the range of 0.07 – 0.50 nm$^{-1}$ with the step size of 0.005 nm$^{-1}$.

The deviation of the unit cell parameters within each sample were estimated from the Bragg peak widths using the Williamson-Hall method (the details of the analysis are given in section S3.5, Supporting Information).[38] The extracted radial superlattice distortion $g_q$ provides the dispersion of the unit cell parameter in a single sample as $\delta a = g_q \cdot \langle a \rangle$, where $\langle a \rangle$ is the average unit cell parameter extracted by the AXCCA. The extracted angular lattice distortion $g_\varphi$ resembles the dispersion in the angular unit cell parameter as $\delta\varphi = g_\varphi \cdot 180°/\pi$. Thus, the extracted dispersion in radial and angular unit cell parameters for all samples is provided in **Table 1** as an error.

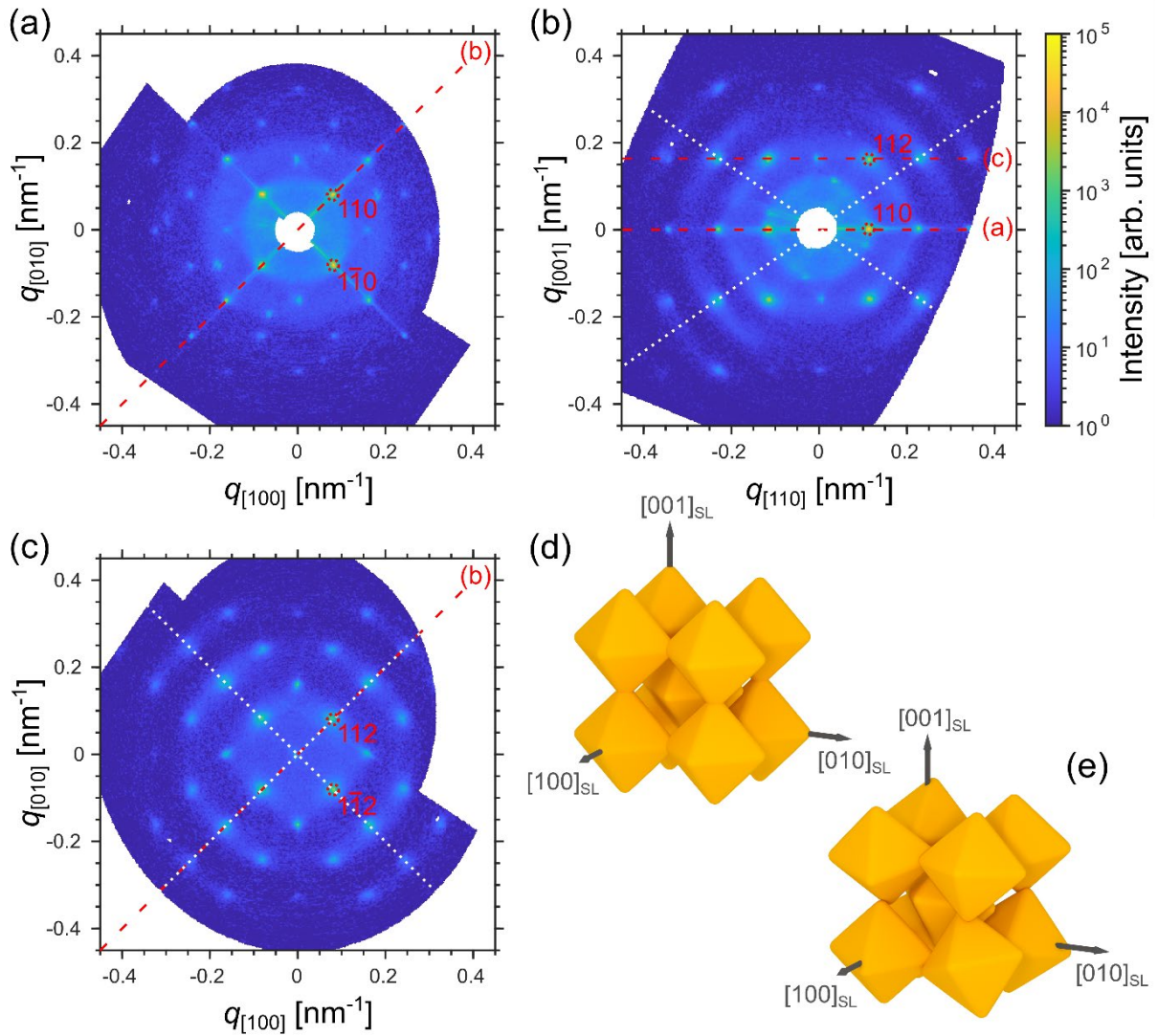

**Figure 5.** a-c) Cuts through the measured 3D intensity distributions for the O1 sample: (a) through the reciprocal space origin and normal to the $[001]_{SL}$ axis, (b) through the reciprocal space origin and normal to the $[1\bar{1}0]_{SL}$ axis, and (c) through the 002 Bragg peak and normal to the $[001]_{SL}$ axis (the offset along the $[001]_{SL}$ axis is 0.162 nm$^{-1}$). The red dashed lines indicate the cuts shown in the other panels. Along with the Bragg peaks, anisotropic features of the form factor of the gold octahedral nanoparticles are well visible. The white dotted lines in panel (b) indicate the $[111]_{SL}$ and $[11\bar{1}]_{SL}$ axes, and in panel (c) – the $[110]_{SL}$ and $[1\bar{1}0]_{SL}$ axes. (d, e) The real space models of a *bcc* superlattice unit cell with the ideal angular orientation of the nanooctahedra (d) and with a certain degree of angular disorder (e). The orientation of the octahedral nanoparticles within the superlattice is revealed by analysis of the anisotropic form factor.

After analysis of the Bragg peaks originating from the superlattice, we turn our attention to the angular orientation of individual octahedra inside the superlattice. The octahedral shape of the nanoparticles is defined by eight $\{111\}_{AL}$ facets that lead to a highly anisotropic form factor of a single nanooctahedron with fringes normal to the facets. The cuts of the 3D intensity distribution through the crystallographic planes of the superlattice, presented for O1 sample in **Figure 5a–c** (the cuts for other samples are given in **Figures S12.1 – S12.4**, Supporting Information), show the presence of the anisotropic form factor features. This finding confirms that the nanooctahedra inside the superlattice are mutually oriented. As can be seen from **Figure 5b**, the fringes normal to the nanooctahedra facets are oriented along the $\langle 111 \rangle_{SL}$ axes. Alignment of the form factor features with the $\langle 110 \rangle_{SL}$ axes in **Figure 5c** confirms that all $\{111\}_{AL}$ nanooctahedron facets are oriented with the $\langle 111 \rangle_{SL}$ superlattice axes. Thus, all $[hkl]_{AL}$ atomic lattice axes are aligned along the corresponding $[hkl]_{SL}$ superlattice axes as shown in **Figure 5d**. Comparison of the experimentally measured scattering patterns with the simulated ones for the given orientation of nanooctahedra (see section S3.6, Supporting Information) confirms our observations. Broadening of the form factor features in the azimuthal direction implies orientational disorder of the nanooctahedra around their ideal orientation as shown in **Figure 5e**. By visual inspection of the measured scattered intensity distribution with the simulated ones for different degree of orientational disorder, we concluded that the nanooctahedra have a normal distribution around the ideal orientation with a standard deviation of $\delta\varphi \approx 15°$.

Although the *bcc* superlattice structure with tip-to-tip oriented nanooctahedra does not provide the highest possible packing density for nanooctahedra, it was previously reported for mesocrystals consisting of nanooctahedra capped by organic ligands.[15] Such a structure is believed to ensure the packing of ligands in the voids formed by the facets of adjacent nanooctahedra. This speculation is supported by the difference in the ligand density on the facets and close to tips.[6] The lower ligand density at the tips facilitate the tip-to-tip orientation of the nanooctahedra in the *bcc* superlattice.

The revealed angular orientation of the nanooctahedra makes it possible to compare the defined superlattice unit cell parameters with the nanooctahedra dimensions. First, we extracted the nanooctahedra dimensions from the TEM measurements. An average edge length $l$ was determined from the projections of individual nanooctahedra along the $[111]_{AL}$ direction for each sample (see section S1.2, Supporting Information). From the edge length, the facet-to-facet distance $d_{f-f}$, and the ideal tip-to-tip distance $d_{t-t}^{ideal}$ were calculated as $d_{f-f} = \sqrt{6}l/3$ and $d_{t-t}^{ideal} = \sqrt{2}l$, respectively. All obtained values are summarized in **Table 1**. We note that the obtained values correspond to dimensions of an ideal nanooctahedron and do not take into account truncation of the nanooctahedra corners, which is visible on the TEM images (see

**Figure S1**, Supporting Information) ) and confirmed by the HAADF-TEM tomography reconstruction (see **Figure 1f**). The facet-to-facet distance $d_{f-f}$ is not affected by the truncation and thus corresponds to the actual size of the nanooctahedra.

As it follows from **Figure 5d**, the nearest neighbor nanooctahedra along the $\langle 111 \rangle_{SL}$ directions face each other with their facets. The nearest-neighbor distance $d_{NN}$ in the *bcc* superlattice can be calculated from the unit cell parameter *a* as $d_{NN} = \sqrt{3}a/2$. The obtained values $d_{NN}$ given in **Table 1** can be compared to the facet-to-facet distance $d_{f-f}$ obtained from the TEM measurements as described above. The difference of these values corresponds to the spacing between the nanooctahedra and is mainly due to the interspacing polymer. As expected, the facet-to-facet distance is smaller than the nearest neighbor distance for all studied samples with the interparticle distance of about 8 nm regardless of the nanooctahedra size.

In its turn, the adjacent nanooctahedra along the $\langle 100 \rangle_{SL}$ axes touch each other with their tips. The unit cell parameters *a*, equal to the distance between the adjacent nanooctahedra in these directions and summarized in **Table 1,** can be directly compared to the tip-to-tip distance $d_{t-t}^{ideal}$ obtained from the TEM measurements. Apparently, the tip-to-tip distances are larger than the unit cell parameters for all the samples; the difference gradually decreases from about 24 nm for the biggest nanooctahedra down to about 7 nm for the smallest ones. For an ideal lattice of untruncated nanooctahedra, this would lead to a topologically unfavorable structure. In the experiment, we deal with nanooctahedra with slightly truncated tips that lead to the reduction of the particle "overlap". Moreover, the observed angular disorder enables more effective packing where the nanooctahedra touch the next ones not directly with their tips but with their edges or facets close to the tips as shown in **Figure 5e**. Thus, the truncation and angular disorder of the nanooctahedra enable the otherwise topologically unfavourable structure.

**Table 1.** The parameters of the samples O1 – O5 as determined from the TEM, AXCCA, and Williamson-Hall method. The edge length *l* was measured from the TEM, error bars are the standard deviation of a few measured samples. The tip-to-tip distance $d_{t-t}^{ideal}$ and facet-to-facet distance $d_{f-f}$ were calculated from the edge length as described in section S1.3 of Supporting Information. The first Bragg peak *q*-value is obtained from the average radial profiles shown in Figure 3c. The unit cell parameter *a* is determined from the AXCCA method and error bars are deduced as the standard deviation obtained from the Williamson-Hall method. The nearest-neighbor distance $d_{NN}$ is obtained from the unit cell parameter *a* as $d_{NN} = \sqrt{3}a/2$. The angular unit cell distortion and error bars are obtained from the Williamson-Hall method.

| SAMPLE | O1 | O2 | O3 | O4 | O5 |
|---|---|---|---|---|---|
| Nanooctahedra edge length $l$ from TEM images [nm] | 72.0 ± 1.6 | 58.4 ± 1.8 | 58.4 ± 1.8 | 46.3 ± 1.4 | 36.7 ± 1.1 |
| Tip-to-tip distance $d_{t-t}^{ideal}$ calculated for ideal nanooctahedron [nm] | 98.5 ± 2.45 | 79.5 ± 2.5 | 79.5 ± 2.5 | 63.8 ± 2.3 | 51.5 ± 1.9 |
| Facet-to-facet distance $d_{f\text{-}f}$ [nm] | 58.8 ± 1.3 | 47.7 ± 1.5 | 47.7 ± 1.5 | 37.8 ± 1.1 | 30.0 ± 0.9 |
| First Bragg peak $q$-value [nm$^{-1}$] | 0.114 | 0.137 | 0.134 | 0.173 | 0.197 |
| Optimized unit cell parameter $a$ [nm] | 77.6 ± 0.9 | 64.5 ± 1.4 | 66.0 ± 1.0 | 51.2 ± 0.5 | 44.8 ± 0.6 |
| Nearest-neighbor distance $d_{NN}$ [nm] | 67.2 ± 0.8 | 55.9 ± 1.2 | 57.2 ± 0.9 | 44.3 ± 0.4 | 38.8 ± 0.5 |
| Angular distortion of the unit cell [°] | 1.0 ± 0.3 | 0.9 ± 0.3 | 0.9 ± 0.4 | 1.3 ± 0.6 | 0.9 ± 0.6 |

## 3. Conclusions

In summary, we presented a synthesis procedure for PS-SH functionalized gold octahedral nanoparticles of variable size (edge length 37, 46, 58, and 72 nm) and investigated the impact of polymer chain length (Mw: 11k, 22k, 43k, and 66k g·mol$^{-1}$) on the growth of faceted colloidal crystals (i.e., mesocrystals) from toluene solution using a gas-phase diffusion technique. To resolve the structure of the grown rhombicdodecahedral mesocrystals we performed an XRD experiment in SAXS geometry at a synchrotron source. By applying the AXCCA method, we identified that in all cases nanocrystals self-assemble in a *bcc* superstructure, where gold octahedral nanoparticles are predominantly oriented tip-to-tip. This, at first, seemingly unfavorable orientational order is possible due to a slight truncation of nanooctahedra along the ⟨100⟩ direction and their angular disorder within the superlattice, which enables a more effective packing. HAADF-STEM tomographic reconstruction revealed the degree of octahedra truncation along ⟨100⟩ directions to be approximately α = 0.14±0.02.

This experimental system provides an important contribution in understanding the packing behavior of octahedral nanoparticles with tunable surface properties (due to the functionalization by polymers with variable chain length) and reveals conditions for crystallization of the faceted mesocrystals. The latter is important for studying of optical properties of such highly ordered assemblies with a tunable distance between the plasmonic nanoparticles (due to surfactants with different polymer chain length) and different coordination geometries. This coordination is determined not only by the symmetry of the superlattice, but also by the location of nanoparticles within the bulk volume, and the orientation of vertices of the faceted mesocrystals. The exact coordination structure of faceted mesocrystals can be further explored in the future by Coherent X-ray Diffraction Imaging (CXDI), which can achieve nanoscale spatial resolution, allowing the presence of different types of structural defects to be revealed.

*4. Experimental Section/Methods*

### *Synthesis and Functionalization of Polystyrene-stabilized Au Nanooctahedra*

The synthesis of the nanocrystals is conducted as published previously,[35] adaptions for the adjustment of the particle morphology were made. The functionalization of the gold nanooctahedra is adapted from a procedure described by Sánchez-Iglesias et al.[36] Four different polystyrene weights are used for this functionalization (11k g·mol$^{-1}$, 22k g·mol$^{-1}$, 43k g·mol$^{-1}$, and 66k g·mol$^{-1}$). The nanoparticle dispersion is centrifuged at 9000 rpm for 5 min, the supernatant is removed and the sedimented particles are redispersed in 10 mL water. The particles are then concentrated to 1 mL (concentration of Au is 2 mg/mL). This dispersion is added to a THF solution containing one ω-thiol-terminated polystyrene chain per nm² surface of the added nanoparticles (estimated from TEM size investigation and assuming ideal octahedral shape) and mixed thoroughly. The dispersion is left undisturbed for 12 h and subsequently centrifuged at 9000 rpm for 5 min, the supernatant is discarded and the particles are redispersed in 10 mL THF. After another round of centrifugation, the supernatant is removed and the particles are redispersed in 1 mL of toluene. This second round of centrifugation is crucial to reduce the amount of water present in the system that would otherwise lead to the irreversible aggregation of the particles upon addition of toluene. The functionalized particles were investigated by TEM (**Figure 1**), UV/Vis (**Figure S1**, Supporting Information), and DLS (**Figure S3**, Supporting Information).

*Self-assembly by gas-phase diffusion*

The nanocubes were assembled to mesocrystals using a gas-phase diffusion technique (**Figure S1** and section S1.4, Supporting Information).[29] 400 µL of 2 mg/mL nanocube dispersion in toluene was filled in a 1 mL flat bottom vial. This small vial was then placed in a larger vial containing an antisolvent consisting of 1 mL ethanol and 1 mL toluene. The larger vial was sealed creating a closed system. The antisolvent diffuses into the nanoparticle dispersion destabilizing the particles and leading to self-assembly indicated by a gradual discoloration of the dispersion. After 3 weeks, the solution was fully discolored. The solution was then overlaid with 200 µL of ethanol to discourage redissolution of the assembled superstructures in the collection process of the self-assemblies. The clear solution was removed using a syringe. The superstructures laying on the bottom were then collected with the residual solvent using a micropipette and placed on a silicon snipped for further investigation. An overview of the self-assembled structures is provided in **Figure 1c**.

*X-ray Scattering Experiment*

The experiment took place at In-situ X-ray diffraction and imaging beamline P23 of the PETRA III storage ring at DESY (Germany). Monochromatic X-rays with a photon energy of 8.32 keV ($\lambda$ = 0.149 nm) were focused by beryllium (Be) compound refractive lenses (CRLs). At this photon energy, the approximate beam size in the vertical direction was 7 – 12 µm (Full Width at Half-Maximum, FWHM) and about 21 – 36 µm (FWHM) in the horizontal direction. This allowed the X-ray beam to completely cover the colloidal crystal grains which have sizes ranging between 5 – 10 μm (compare with **Figure S5**, Supporting Information). The mesocrystal grains were mounted on a tungsten tip and rotated around the vertical axis in steps of 0.5° in the range of 0 – 180°. For each angular position, 2D far-field scattering patterns were collected by the X-Spectrum Lambda 750K GaAs detector, which was positioned 1.497 m downstream from the sample. Series of two frames of 1 s exposure each were measured at each angular position, resulting in a total of 2 s exposure. The sample was cooled by a liquid nitrogen cryojet (OXFORD Cryostream 700) to minimize radiation damage to the organic ligands (stabilizing nanocrystals), which might lead to nanoparticle coalescence and the loss of superlattice organization. The collected 2D patterns were interpolated onto a 3D orthonormal grid with the voxel size of 0.0015 nm$^{-1}$.

*Angular X-ray Cross-Correlation Analysis*

We used the AXCCA method modified for applications to 3D intensity distributions as described in Ref.[37]. The method is based on analysis of two-point CCFs $C(q_1,q_2,\Delta)$ calculated for measured scattered intensity distribution in 3D reciprocal space as

$$C(q_1, q_2, \Delta) = \langle \tilde{I}(\mathbf{q_1})\tilde{I}(\mathbf{q_2})\delta(\frac{\mathbf{q_1} \cdot \mathbf{q_2}}{\|\mathbf{q_1}\|\|\mathbf{q_2}\|} - \cos \Delta)\rangle, \quad (1)$$

where $\tilde{I}(\mathbf{q_1})$ and $\tilde{I}(\mathbf{q_2})$ are normalized intensities taken at the momentum transfer positions $\mathbf{q_1}$ and $\mathbf{q_2}$ in 3D reciprocal space and $\delta(x)$ is a Dirac delta-function. The averaging is performed over the all momentum transfer positions of $\mathbf{q_1}$ and $\mathbf{q_2}$ with the lengths $q_1 = \|\mathbf{q_1}\|$ and $q_2 = \|\mathbf{q_2}\|$, respectively. The intensities are normalized to the mean values at the corresponding $q$-value as

$$\tilde{I}(\mathbf{q}_i) = \frac{I(\mathbf{q}_i) - \langle I(\mathbf{q}_i)\rangle}{\langle I(\mathbf{q}_i)\rangle}, i = 1,2 \quad (2)$$

where averaging is performed over all positions corresponding to $\mathbf{q}_i$ with the length $q_i = \|\mathbf{q}_i\|$.

In this work, we calculated CCFs $C(q_1,q_2,\Delta)$ for a fixed momentum transfer value $q_1$ corresponding to the first Bragg peak position and varying momentum transfer values $q_2$. These CCFs are represented as corresponding maps in $(q_2,\Delta)$-coordinates.

**Supporting Information**

Supporting Information is available from the Wiley Online Library or the corresponding authors.

*Acknowledgments*

We acknowledge DESY (Hamburg, Germany), a member of the Helmholtz Association HGF, for the provision of experimental facilities. Parts of this research were carried out at PETRA III P23 "In situ X-ray diffraction and imaging beamline" and we would like to thank P23 beamline engineer Atula Poduval for assistance during the beamtime. Beamtime was allocated for proposal I-20210458 . The authors want to acknowledge the support of the core facilities nano.lab (Nanostructure Laboratory) and PAC (Particle Analysis Centre) of the University of Konstanz. F. K. and E. V. S acknowledge the DFG (Deutsche Forschungsgemeinschaft) for financial support (SFB 1214. Project B1).

**Conflict of interest**

The authors declare no conflict of interest.

**Data availability**

The X-ray data that support the findings of this study are openly available in Zenodo.org at https://zenodo.org/record/8060535

**Author contributions**

D. L. and S. S. contributed equally to this work. F.K. synthesized the nanoparticles and prepared the samples. S. St., Th. W., Th.G., E.V.S. and A.L. prepared the samples for X-ray measurements. S. St., E.V.S. and K. M-C. performed HAADF-STEM measurements, D. L., S. S., D. A., A. I., A. Kh., D. N., and I.A.V. performed the X-ray scattering experiment. D. L. and S. S. analyzed the collected X-ray data. E.V.S. and I.A.V. conceived and supervised the project. D. L., S. S., F. K., S. St., E. V. S. and I. A. V. wrote the manuscript with input from all authors. All authors have given approval to the final version of the manuscript.

**Supporting Information**

## S1. NANOPARTICLE SYNTHESIS AND SELF-ASSEMBLY

The synthesis of the nanostructured material included the synthesis of nanooctahedra in aqueous solution, their phase transfer to toluene combined with a functionalization with polystyrene. This particle dispersion was subsequently used to assemble superstructures using a gas-phase diffusion technique. The particles and superstructures were characterized using TEM, SEM, UV/Vis, DLS, and X-ray scattering (see Section S3)

**S1.1 Chemicals and Methods**

Milli-Q water (resistivity 18.2 MΩ·cm) was used in all experiments. If not mentioned otherwise, all chemicals were bought and used as received. Hydrogen tetrachloroaurate trihydrate (HAuCl$_4$ · 3 H$_2$O, ≥ 99.9 %), hexadecyltrimethylammonium bromide (CTAB, ≥ 99 %), cetyltrimethylammonium chloride solution (CTAC, 25 wt. % in H$_2$O) and hexadecylpyridinium chloride monohydrate (CPC, 99.0 – 102.0 %) were purchased from Sigma Aldrich. Sodium borohydride (NaBH$_4$, ≥ 97 %) and L(+)-ascorbic acid (AA, ≥ 99 %) were purchased from Roth. Potassium bromide (KBr, for IR spectroscopy) was purchased from Merck. Poly(styrene), ω-thiol-terminated polystyrene was purchased from Polymer Source Inc. THF (≥ 99.5 %, stabilized), toluene (≥ 99.5 %), and ethanol (absolute ≥ 99,8 %) was purchased from VWR.

UV-Vis spectra were recorded using an Agilent Cary 60 UV-Vis Spectrophotometer. Dynamic light scattering was measured on a Malvern Zetasizer Nano ZSP. SEM was performed on a Zeiss Gemini 500 operating at 2 kV acceleration voltage and a Zeiss 1540XB CrossBeam FIB SEM. For TEM studies, NPs were deposited on Quantifoil copper grids with 2 nm-thick carbon films. TEM imaging was conducted on a Zeiss Libra120 with lanthanum hexaboride emitter operated at 80 kV and equipped with a Koehler illumination system. HAADF-STEM-Tomography has been performed at a FEI Themis operated at 300kV.

**S1.2 Synthesis of nanoparticles**

The synthesis of the nanoparticles was conducted as published previously,[1] and adaptions for the adjustment of the particle morphology were made.

<u>Synthesis of initial seeds.</u> An aqueous solution of $HAuCl_4$ (0.25 mL, 0.01 M) and CTAB (7.50 mL, 0.10 M) were mixed in a glass vial (50 mL) and tempered to 27 °C. 0.60 mL of freshly prepared, ice-cold solution of $NaBH_4$ (0.01 M) was added under vigorous stirring. The resulting solution turned brown immediately. The seed solution was aged for 90 min at 27 °C to ensure complete decomposition of excess borohydride.

<u>Synthesis of spherical seeds.</u> The synthesis of spherical seeds was adapted from Zheng et al.[2] Aqueous solutions of CTAC (39.00 mL, 0.10 M) and $HAuCl_4$ (1.00 mL, 0.01 M) were mixed in a glass vial (100 mL) and tempered to 27 °C. AA (15.00 mL, 0.10 M) was added, followed by rapid injection of 500 µL initial seeds. The solution turned red immediately and was kept at 27 °C for 15 min. The spherical seeds were collected by centrifugation at 30000 RPM for 120 min, washed once with water and twice with CPC solution (0.02 M). The optical density of the dispersion was adjusted to 0.275 (approx. $1 \times 10^{12}$ particles·ml$^{-1}$).

<u>Synthesis of faceted nanoparticles.</u> Aqueous solutions of $HAuCl_4$ (1 ml, 0.01 M), KBr, and CPC (50 ml, 0.1 M) were mixed in a glass vial (100 ml) and tempered to 27 °C. An aqueous solution of AA (1.5 mL, 0.1 M) was added, followed by rapid injection of washed spherical seeds. The solution turned pinkish red and was kept at 27 °C for 3 h. The NPs were collected by centrifugation at 9000 RPM for 5 min and redispersed in CPC solution (0.02 M). For discrete amounts of KBr and seeds and the resulting edge lengths *l* of the octahedral see **Table S1**. The TEM images were analyzed by DigitalMicrograph Gatan Microscopy Suite 3 software (Gatan Inc., ver. 3.41.2938.1) and Fiji.[3]

**Table S1.** Amounts of KBr and initial seeds added to the synthesis mixture in relation to the obtained edge length of the octahedra.

| Sample | KBr [mL] | Initial seeds [μL] | Edge length $l\pm\sigma$ [nm] |
|---|---|---|---|
| Size 1 | 4.25 | 400 | 72.0±1.6 |
| Size 2 | 3.75 | 800 | 58.4±1.8 |
| Size 3 | 2.75 | 1600 | 46.3±1.4 |
| Size 4 | 2.25 | 3200 | 36.7±1.1 |

**S1.3 Conventional TEM investigation of nanooctahedra size**

The octahedra sizes are estimated from TEM micrographs assuming ideal octahedral shape of the particles lying on their {111} facets (**Table S1, Figure 1a**). The edges are only slightly truncated (compare also Ref. [1]) allowing for a quite reliable measurement of the edge length $l$ (distance between the sides of the 2D projection along the [111] direction) and calculation of the distance between two opposite {111} facets as $d_{f-f} = \sqrt{6}l/3$. The tip-to-tip distance $d_{t-t}$ can be calculated from $d_{t-t} = \sqrt{2}\, l$. This slightly overestimates the tip-to-tip distance due to the truncation of the synthesized nanooctahedra. In **Figure S1**, a TEM micrograph of a synthesized nanoparticle (a) is compared with a sketched ideal (gray) and truncated octahedra (green) projected along the (b) [111], (b) [110] and (d) [102] axes.

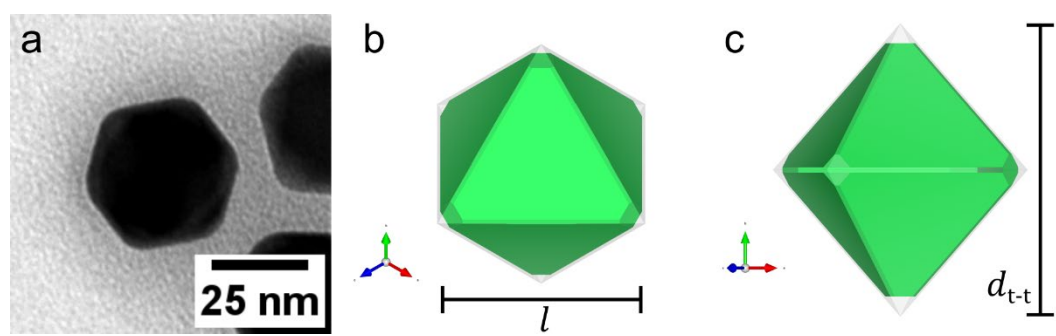

**Figure S1**. The TEM image of an as synthesized nanooctahedron (a) is compared with a sketched ideal (gray) and truncated octahedra (green) projected along [111] (b), and [102] (c) axes. It can be observed that the edge length $l$ may be measured from TEM reliably. However, the calculation of the tip-to-tip distance $d_{t-t}$ using the $l$ leads to an overestimation of this distance due to the tip truncation of the synthesized nanooctahedra. The sketched octahedra are depicted with the truncation of α=0.14, as calculated in Section 2.1.

### S1.4 HAADF-STEM-Tomography investigation of polystyrene-functionalized nanooctahedra

To verify the shape of the individual nanooctahedra used as building blocks for the mesocrystals, HAADF-STEM-tomography has been conducted at the FEI Titan operated at 300 kV with a semi-convergence angle of 23.3mrad and at a camera length of 195 mm, leading to detector acceptance angles from 35 to 200mrad. To this end, nanoparticles were transferred directly onto the TEM-grids from solution and several nanocrystals were selected, of which a tilt series of HAADF images in a tilt range from -72.5° to +75° in 2.5° steps were recorded, aligned and tomographically reconstructed using a weighted simultaneous iterative reconstruction technique (W-SIRT) algorithm.[4]

**Figure S2** (a,b,c) shows three reconstructed nanooctahedra (sample O1), with (100) and (110) planes slicing through the center of nanocrystals. The corresponding slices are shown in **Figure S2** (d,e,f) for the <100> directions and **Figure S2** (h,i,j) for the <110> directions, respectively. For the <100> directions, the edges of an ideal square cross-section with 72 nm side length are overlaid in red and edges of ideal parallelograms for the <110> directions. The degree of truncation along the <100> and <110> directions was estimated by comparison with an ideal octahedra (see Section 2.1.).

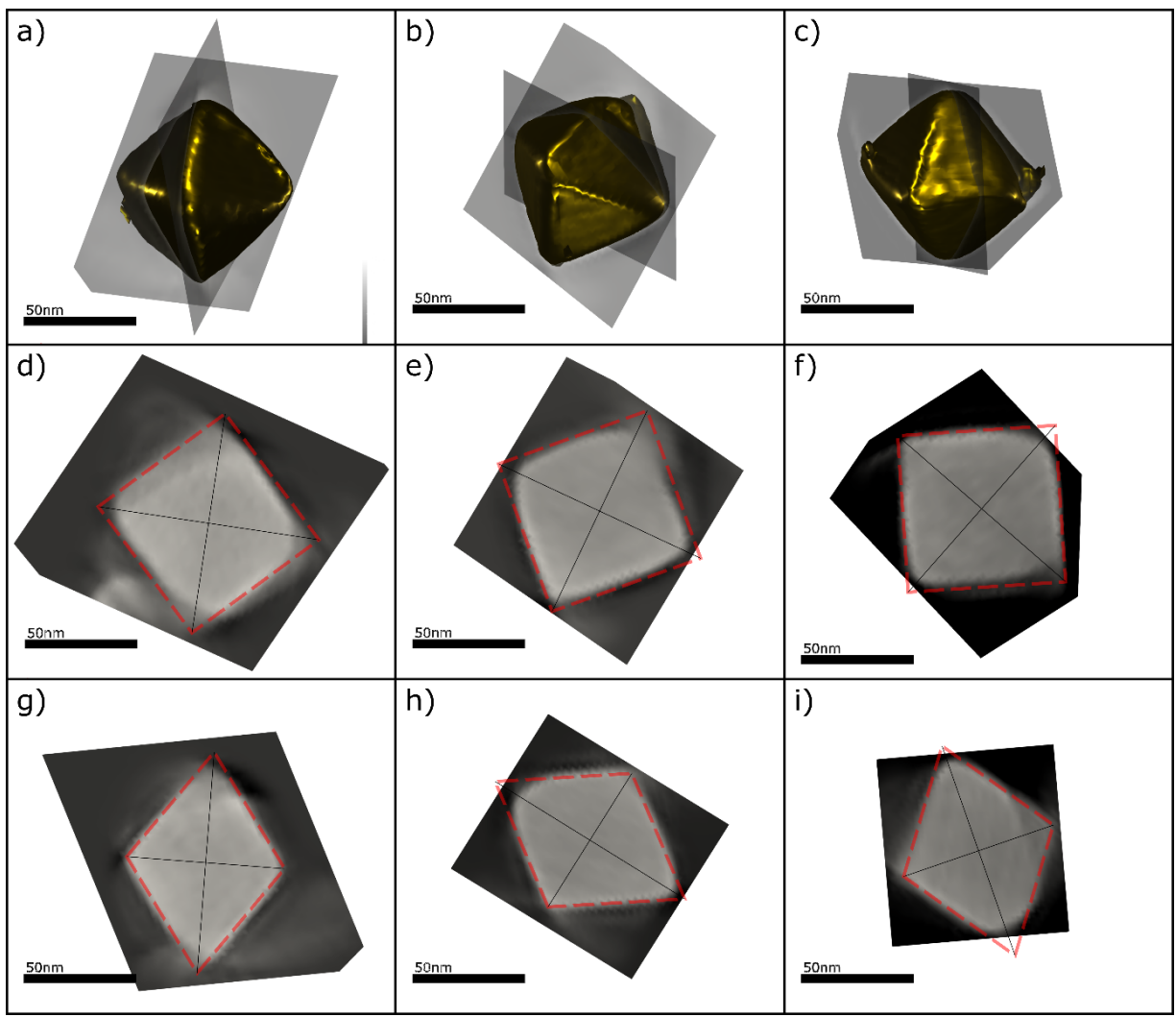

**Figure S2.** HAADF-STEM-Tomographic reconstruction of gold nanocrystals (O1, edge length 72 nm). (a-c) 3D reconstructed nanooctahedra, indicating (100) and (110) planes slicing through the center of nanocrystals. The corresponding slices are shown in (d,e,f) for the <100> directions and in (h,i,j ) for the <110> directions, respectively. The slices are overlayed by an ideal cross-section of an octahedra (red dashed-lines).

**S1.5 DLS investigation of polystyrene-functionalized nanooctahedra**

DLS was measured at 25 °C. As only one particle species and thus one main peak was expected, the Z-Average and polydispersity index (PDI) were selected for evaluation. The Z-average represents the hydrodynamic diameter of the nanoparticles, while the PDI is a scale for the dispersity of the system. In PDI values smaller than 0.05 are mostly measured from monodisperse standards, and values greater than 0.7 indicate that the sample has a very broad size distribution.[5] As the DLS evaluation approximates spherical particles, the sphere equivalent of the nanocrystal cores was calculated based on measurements from TEM imaging. This diameter was then subtracted from the overall hydrodynamic diameter, and the result was halved for an estimation of the extension length of the polymer from the nanoparticle surface into the solution (**Figure S4**).

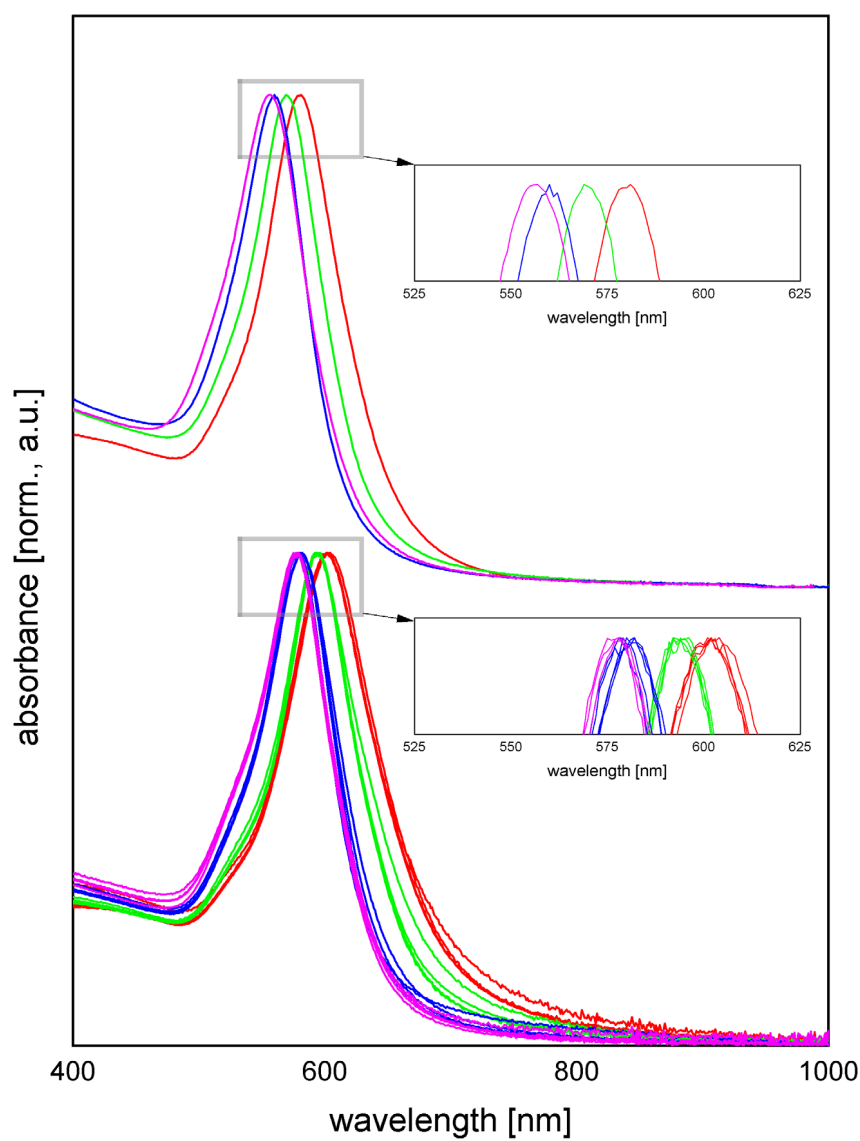

**Figure S3**. UV/Vis spectra (Size 4 magenta, Size 3 blue, Size 2 green, Size 1 red) from CPC-stabilized particles in aqueous dispersion (top) and polystyrene-stabilized in toluene (bottom). Upon functionalization and transfer to toluene, an expected red-shift (see inlay for enlarged spectra) of the absorbance peak is observable while no peak broadening or peak splitting is indicating destabilization or fusion of the dispersed particles. The polymer weight does not significantly influence the peak position, thus the spectra from one size but different polymer weights are combined.

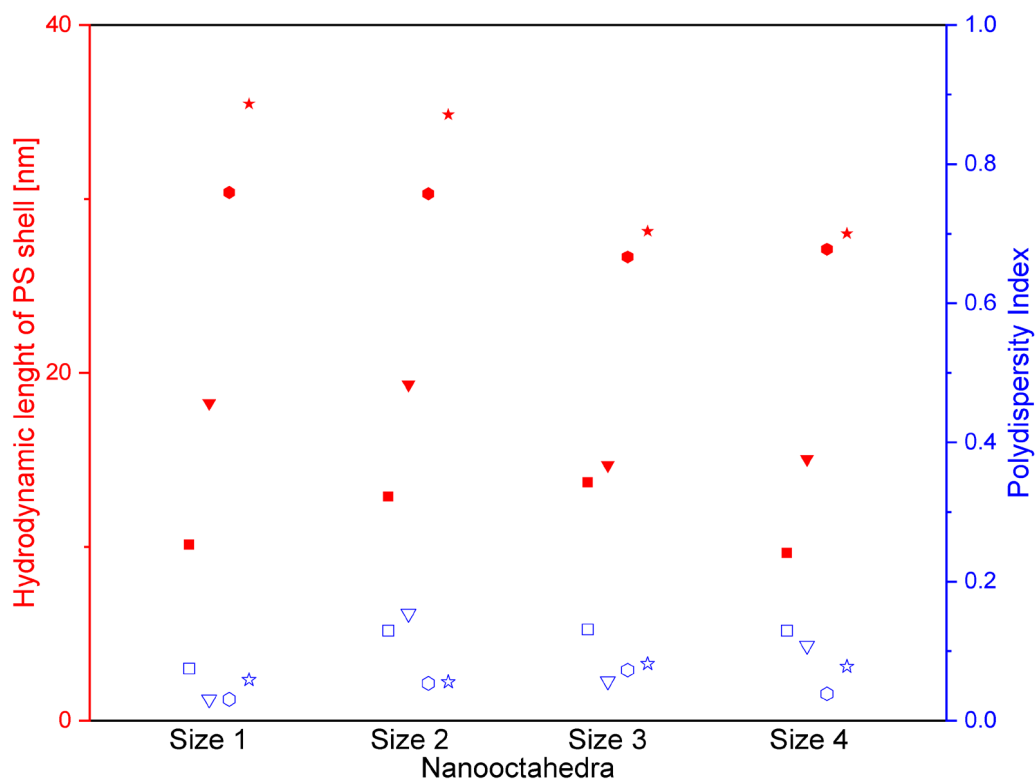

**Figure S4.** DLS measurements reveal a constant rise of the hydrodynamic radii of the nanooctahedra-ligand composites with increasing polymer weight. The hydrodynamic radius of the ligand shell (cube 11k g·mol$^{-1}$, triangle 22k g·mol$^{-1}$, hexagon 43k g·mol$^{-1}$, star 66k g·mol$^{-1}$) is calculated as the difference of the spherical equivalent of the octahedra as measured from the TEM images (sphere diameter: Size 1 70 nm; Size 2 58 nm; Size 3 45 nm; Size 4 35 nm) and the Z-average mean size as evaluated through dynamic light scattering. The polydispersity index proves a quite monodisperse sample composition.

**S1.6 Self-assembly to mesocrystals via a gas-phase diffusion technique**

A sketch of the gas-phase diffusion setup can be found in **Figure 1b** in the main text. This method was already successfully applied for the self-assembly of significantly smaller nanoparticles such as magnetite, lead sulfide, and platinum.[6-9] Toluene was chosen as the dispersing agent for the investigated system due to its low vapor pressure which minimizes evaporation and avoids a drying of the dispersion instead of destabilization by reduction of solvent quality. To further circumvent such drying the antisolvent was composed of ethanol and toluene in a 1:1 volume ratio so that the atmosphere in the vial could be saturated from non-dispersing toluene. Upon diffusion of the antisolvent mixture into the dispersion, the self-assembly of the nanoparticles was indicated by a gradual discoloration of the dispersion. Upon full discoloration of the dispersion (2 – 3 weeks) the assembly process was deemed complete and the colloidal assemblies could be collected from the bottom of the glass vial.

## S2. SAMPLE PREPARATION FOR X-RAY SCATTERING EXPERIMENT

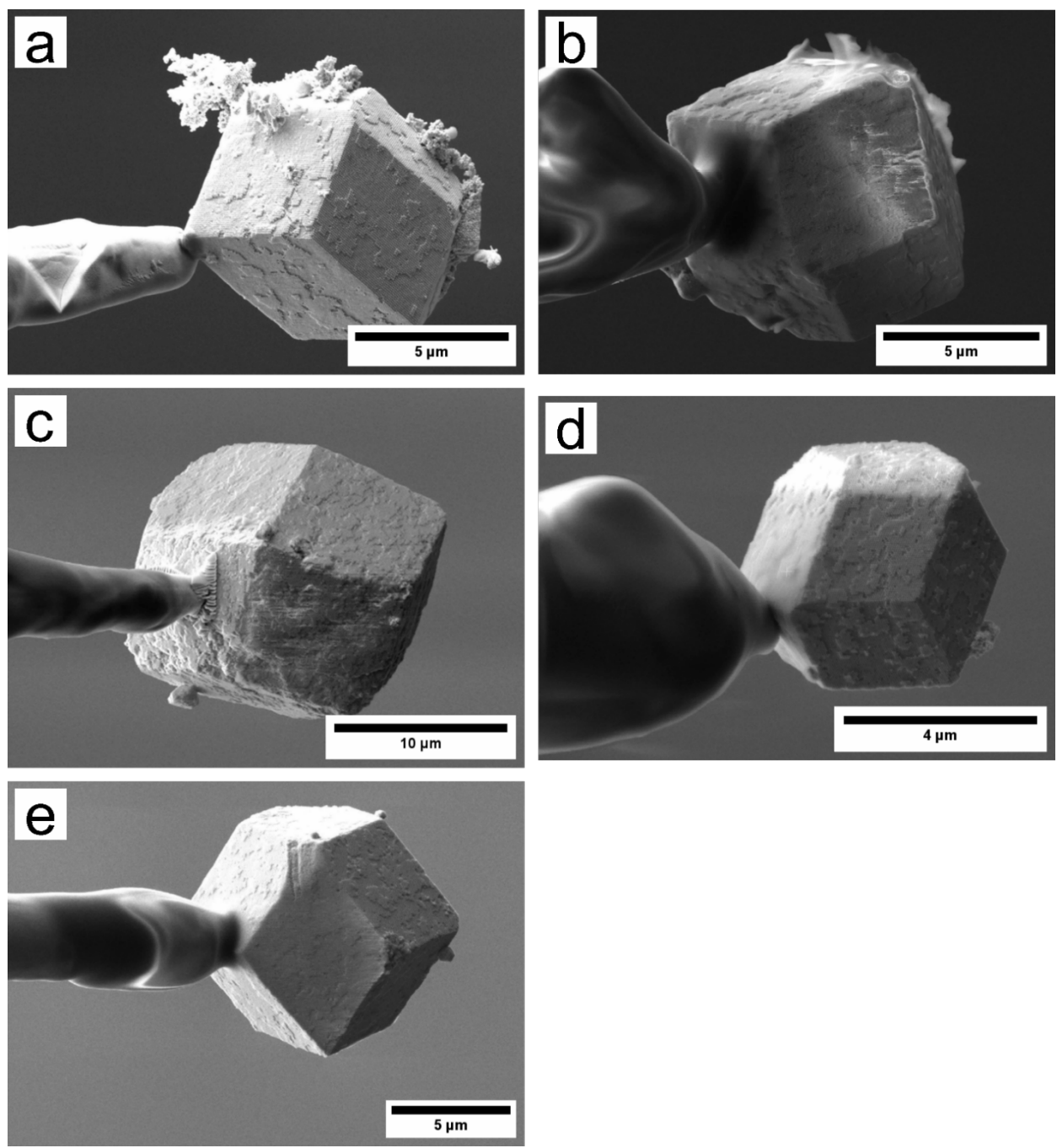

**Figure S5.** Individual mesocrystals were separated and attached to a tungsten needle tip for X-Ray diffraction measurements. SEM images of 5 colloidal crystal samples that were selected and prepared for measurement at the P23 beamline are depicted. a) O1 sample (72 nm edge length, 66k g·mol$^{-1}$ PS-SH). b) O2 sample (58 nm edge length, 66k g·mol$^{-1}$ PS-SH). c) O3 sample (58 nm edge length, 43k g·mol$^{-1}$ PS-SH). d) O4 sample (46 nm edge length, 22k g·mol$^{-1}$ PS-SH). e) O5 sample (37 nm edge length, 22k g·mol$^{-1}$ PS-SH).

# S1. ANALYSIS OF X-RAY SCATTERING DATA

## S1.1 Data pre-processing

The 2D scattering pattern measurement routine is described in the Methods section of the main text. Background scattering patterns without the sample were measured and subtracted from all 2D scattering patterns. The obtained patterns were interpolated onto a 3D orthonormal grid with the voxel size of 0.0015 nm$^{-1}$ by bilinear interpolation. The isosurfaces of the resulting 3D intensity distributions for all measured samples are shown in **Figure S6**.

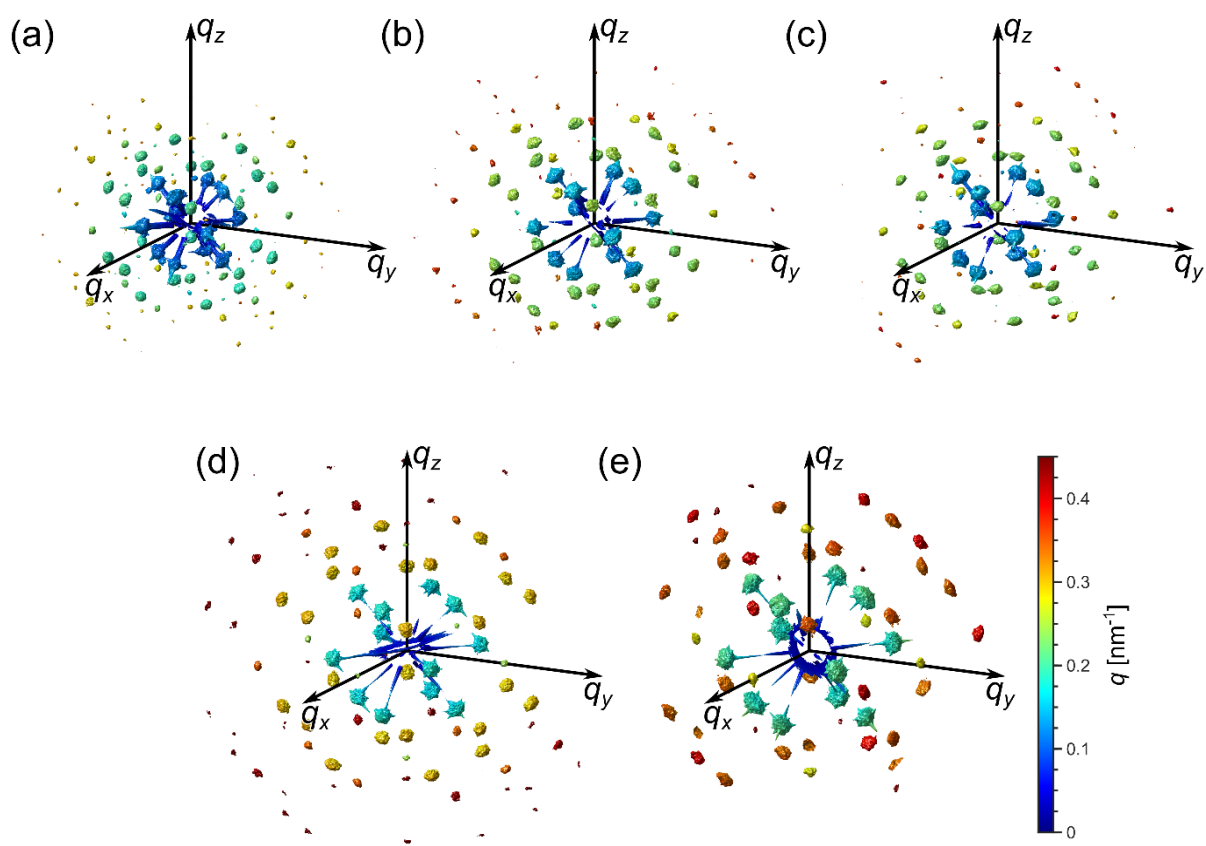

**Figure S6.** Isosurfaces of the obtained 3D intensity distributions in reciprocal space taken at the same isovalue for O1 (a), O2 (b), O3 (c), O4 (d) and O5 (e) samples. The colors represent the $q$-value (distance from the origin of reciprocal space).

WILEY-VCH**S1.2 Radial profile analysis and AXCCA of 3D scattered intensity distributions**

The averaged radial profiles of the scattered intensity $I(q)$ were calculated according to the following equation

$$I(q) = \frac{\sum_{|\|\boldsymbol{q}_i\|-q|<\varepsilon} I(\boldsymbol{q}_i)}{\sum_{|\|\boldsymbol{q}_i\|-q|<\varepsilon} 1} \quad \textbf{(S1)}$$

where $I(\boldsymbol{q}_i)$ is the intensity value at the point of 3D reciprocal space with the coordinates $\boldsymbol{q}_i$ and $\varepsilon$ is the averaging window defining the resolution. In this work we used the value of $\varepsilon = 0.005$ nm$^{-1}$ providing sufficient resolution. The form-factor radial profiles were calculated according to the same **Equation (S1)**, but the points $\boldsymbol{q}_i$ in the vicinity of the Bragg peaks were excluded from the averaging. The resulting radial profiles are shown in **Figure 3c** of the main text.

To reveal the superlattice structure of the samples, we applied the AXCCA method to the obtained 3D scattered intensity distributions. We used the AXCCA method modified for applications to 3D intensity distributions, details of which can be found in Ref.[10] A CCF $C(q_1,q_2,\Delta)$ for the data defined on a grid can be calculated as follows

$$C(q_1, q_2, \Delta) = \frac{\sum_{\{|\|\mathbf{q}_i\|-q_1|<\varepsilon\} \cap \{|\|\mathbf{q}_j\|-q_2|<\varepsilon\} \cap \{\Delta_{ij} \in [\Delta-d\Delta;\Delta+d\Delta]\}} \tilde{I}(\mathbf{q}_i)\tilde{I}(\mathbf{q}_j)}{\sum_{\{|\|\mathbf{q}_i\|-q_1|<\varepsilon\} \cap \{|\|\mathbf{q}_j\|-q_2|<\varepsilon\} \cap \{\Delta_{ij} \in [\Delta-d\Delta;\Delta+d\Delta]\}} 1}, \quad \textbf{(S2)}$$

where $\mathbf{q}_i$, $\mathbf{q}_j$ are the points close to the spheres of the radii $q_1$ and $q_2$ in reciprocal space, respectively, $\Delta_{ij}$ is the relative angle between these points. The sum is calculated over all pairs of points $\mathbf{q}_i$, $\mathbf{q}_j$ with the corresponding relative angle $\Delta$. The angular resolution of $\Delta$ was experimentally set to 0.5° that allows one to resolve all peaks in the resulting CCFs. The angular averaging window $d\Delta$ was correspondingly set to 0.25°. The value of the first momentum transfer $q_1$ was fixed to the position of the first Bragg peak in the radial profiles shown in **Figure 3c** in the main text (see **Table 1** in the main text for the values) and the value of the second momentum transfer $q_2$ was varied in the range of $q_2 = 0.07 - 0.50$ nm$^{-1}$ with a step size of 0.005 nm$^{-1}$. The radial averaging window $\varepsilon$ was correspondingly set to 0.0025 nm$^{-1}$. The resulting CCFs for all samples are shown in ($q_2$,$\Delta$)-coordinates in **Figure S7**.

In order to determine the lattice symmetry and parameter, modelling of the peak positions was performed as described in Ref.[10] The model unit cell was defined by three basis vectors in real space. The corresponding reciprocal basis vectors were calculated and used to calculated the expected peak positions in the CCFs for all possible reciprocal lattice points corresponding to the Bragg peaks in the scattered intensity. All samples were found to have a *bcc* structure. The corresponding unit cell parameter *a* was varied to fit the peak positions in the experimentally obtained CCFs. The expected peak positions for the optimized unit cell parameters are shown in **Figure S7** with red circles. The coincidence of these positions with the peaks present in the CCFs made it possible to determine the unit cell parameters summarized in **Table 1** in the main text.

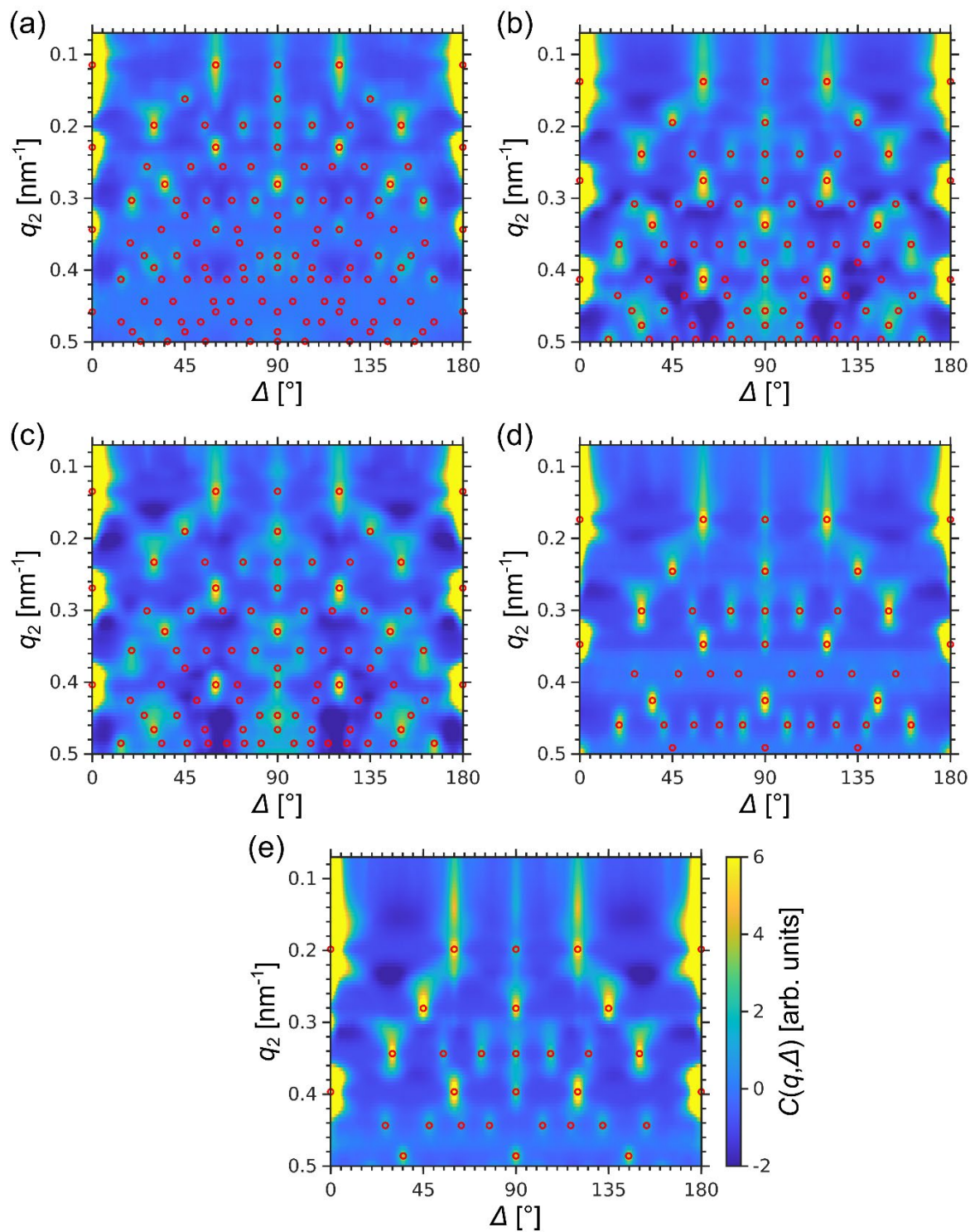

**Figure S7.** (a-e) 2D maps of cross-correlations functions $C(q_1, q_2, \Delta)$ calculated for samples O1 (a), O2 (b), O3 (c), O4 (d) and O5 (b). Red circles represent the peak positions for the optimized bcc structure. The cross-correlation functions calculated for $q_1$ correspond to the first Bragg peak position (see Table 1 for the values) and $q_2$ was varying in the range of 0.07 – 0.50 nm$^{-1}$ with the step size of 0.005 nm$^{-1}$.

**S1.3 The Williamson-Hall analysis**

The width of the Bragg peaks for each sample were studied by the Williamson-Hall method.[11] We extracted three 1D profiles for each of the Bragg peak in the measured 3D intensity distribution. One profile represents the radial intensity profile, two other profiles were extracted in two orthogonal angular directions as shown in **Figure S8**.

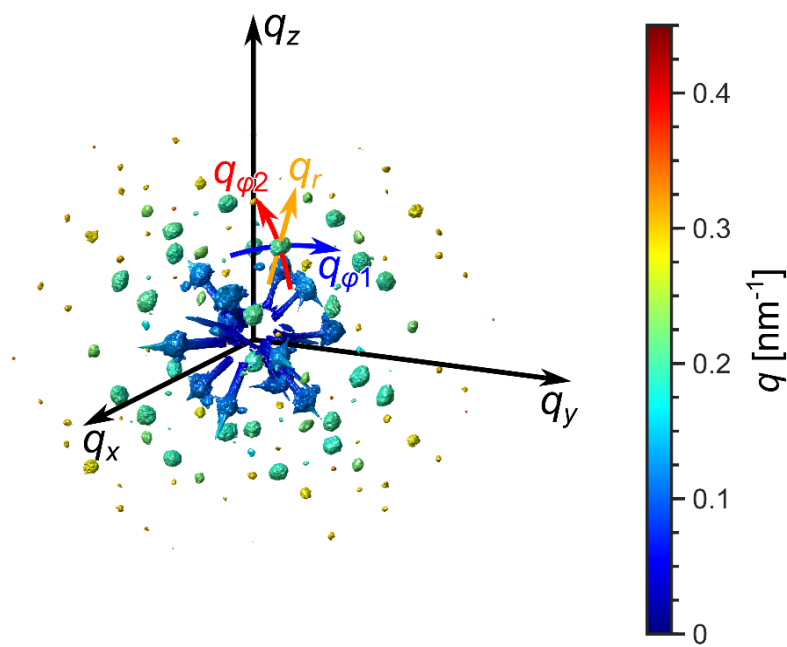

**Figure S8.** Isosurface of the obtained 3D intensity distribution in reciprocal space for O1 sample. The colors represent the *q*-values (distance from the origin of reciprocal space). The orange line indicates the radial direction, the red and blue lines indicate two orthogonal angular directions along which the intensity profiles of a single Bragg peak were analyzed.

The corresponding widths were extracted by fitting the 1D profiles $I(q)$ of each Bragg peak separately with the Lorentzian function

$$I(q) = \frac{I_0 \sigma_{q/\varphi}}{\pi} \frac{1}{(q-q_0)^2 + \sigma_{q/\varphi}^2}, \qquad \text{(S3)}$$

where $I_0$ is the integrated intensity, $q_0$ is the peak center coordinate, $\sigma_{q/\varphi}$ is the peak half-width at half-maximum value in radial ($q$) and angular ($\varphi$) directions.

The extracted widths were averaged for each family of the Bragg peaks and plotted against the *q*-values as shown in **Figure S9**. The error bars represent the standard deviation within each Bragg peak family. We did not observe any substantial difference in the width in two orthogonal angular directions that indicates the isotropic character of broadening.

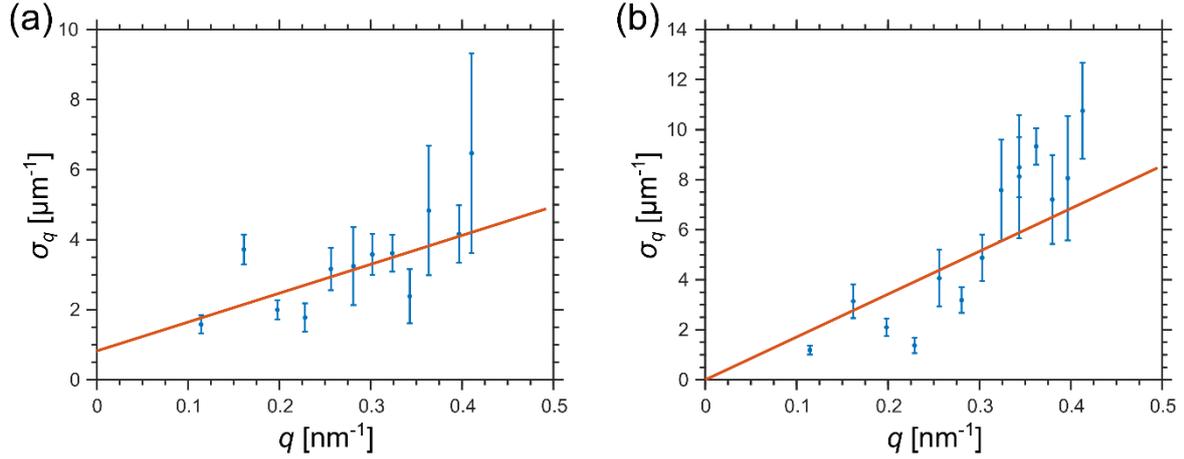

**Figure S9.** An example of the Williamson-Hall plots for the radial (a) and angular (b) widths of the Bragg peaks for O1 sample. Blue points represent the values extracted from the experimentally measured intensity distribution, and the red lines represent the best fits with **Equation (S4)**. The error bars represent the standard deviation within each Bragg peak family.

The dependence of the Bragg peak width on the *q*-values is reasoned by two factors: the Scherrer's broadening due to the finite coherently scattering volume and the broadening due to the unit cell parameter distribution. For the Lorentzian shape of the peaks, this dependence can be described by the Williamson-Hall equation

$$\sigma_{q/\varphi}(q) = \frac{2\pi}{D_{q/\varphi}} + g_{q/\varphi} q, \qquad (S4)$$

where $\sigma_{q/\varphi}$ is the half-width at half-maximum value, $D_{q/\varphi}$ is the coherently scattering domain size and $g_{q/\varphi}$ is the lattice distortion value. The indices $q/\varphi$ denote the radial and angular directions, respectively. The radial lattice distortion value $g_q$ is defined as $g_q = \sigma_a/\langle a \rangle$, where $\langle a \rangle$ is the mean unit cell parameter and $\sigma_a$ is the half-width at half-maximum of the unit cell parameter distribution. The extracted $\sigma_a$ value is interpreted as the standard deviation of the unit cell parameters represented in **Table 1** in the main text. The angular lattice distortion value $g_\varphi$ is equal to the half-width at half-maximum $\sigma_\varphi$ of the unit cell angular parameter distribution $g_\varphi = \sigma_\varphi$. The extracted $\sigma_\varphi$ values are given in **Table 1** in the main text.

Apparently, the disorder is the main reason for the Bragg peak broadening for these samples (the offset in the Williamson-Hall plots is close to zero, see **Figure S9**). This fact, together with the large error bars, makes impossible estimation of the coherently scattering domain size.

**S1.4 Nanooctahedra orientation determination**

The octahedral shape is defined by eight $\{111\}_{AL}$ facets that leads to a highly anisotropic form factor of a single nanooctahedron with fringes normal to the facets. This anisotropy allows determination of the angular orientation of the nanooctahedra in the superlattice. The slices normal to the $[001]$ and $[1\bar{1}0]$ axes shown in **Figure 5** of the main text are the most illustrative ones. From this figure, it is clear that all facets are perpendicular to the $\{110\}$ planes that implies collinearity of all crystallographic directions of the nanoparticle atomic lattice to those of the superlattice $[hkl]_{AL} \parallel [hkl]_{SL}$ for any Miller indices $h$, $k$, $l$.

To prove that, we compared the slices of the measured scattered intensity with cuts of the simulated scattered intensity from an ensemble of nanooctahedra. The form factor amplitude of an ideal octahedron can be calculated according to the following equation[12]

$$F(\mathbf{q}, R) = \frac{6}{R}\left[\frac{q_y \sin(q_y R) - q_z \sin(q_z R)}{(q_x^2 - q_y^2)(q_y^2 - q_z^2)} + \frac{q_x \sin(q_x R) - q_z \sin(q_z R)}{(q_x^2 - q_z^2)(q_y^2 - q_x^2)}\right], \quad \text{(S5)}$$

where $R$ is half of the tip-to-tip distance ($R = l/\sqrt{2}$, where $l$ is the octahedron edge length). The coordinates are chosen in such a way that the nanooctahedron have the same orientation as shown in **Figure 5d** of the main text. The scattered intensity from an ensemble of nanooctahedra with a normal size distribution was calculated as follows:

$$I(\mathbf{q}, \langle R \rangle, \sigma_R) = \int_\mathbb{R} D(R, \langle R \rangle, \sigma_R)|F(\mathbf{q}, R)|^2 dR, \quad \text{(S6)}$$

where $F(\mathbf{q}, R)$ is the form factor amplitude of nanooctahedra with the size $R$ from **Equation (S5)** and $D(R, \langle R \rangle, \sigma_R)$ is the probability density function of a normal distribution with the mean value $\langle R \rangle$ and the standard deviation $\sigma_R$

$$D(R, \langle R \rangle, \sigma_R) = \frac{1}{\sqrt{2\pi\sigma_R^2}} \exp\left[-\frac{(R - \langle R \rangle)^2}{2\sigma_R^2}\right]. \tag{S7}$$

A slice normal to the $[1\bar{1}0]$ axis of the resulting simulated 3D scattered intensity for the nanooctahedra with the edge length $l = 72.0 \pm 1.6$ nm is shown in **Figure S10b**. The simulated pattern contains well-defined fringes diverging from the origin along the $\langle 111 \rangle$ axes, but their shapes does not completely resemble the observed one in experimental data sets.

Among many other factors affecting the scattered intensity, there are two major ones: the angular disorder of the nanooctahedra around their average orientation and the tip truncation observed in TEM measurements (see section S1.3). We considered the first effect by introduction of averaging over an ensemble of nanooctahedra having normal angular distribution around all three degree of freedom as follows:

$$\begin{aligned} I(\mathbf{q}, \langle R \rangle, \sigma_R, \delta\varphi) = \\ = \int_{\mathbb{R}} D(\alpha, 0, \delta\varphi) D(\beta, 0, \delta\varphi) D(\gamma, 0, \delta\varphi) I\big(\mathbf{R_z}(\gamma)\mathbf{R_y}(\beta)\mathbf{R_x}(\alpha)\mathbf{q}, \langle R \rangle, \sigma_R\big) \mathrm{d}\alpha \mathrm{d}\beta \mathrm{d}\gamma, \end{aligned} \tag{S8}$$

where $I(\mathbf{q}, \langle R \rangle, \sigma_R)$ is the scattered intensity distribution from **Equation (S6)**, $D(\alpha, 0, \delta\varphi)$ $D(\beta, 0, \delta\varphi)$ and $D(\gamma, 0, \delta\varphi)$ are the probability density distrubutions of angles $\alpha$, $\beta$ and $\gamma$, respectively, around the zero value with the standard deviation $\delta\varphi$ from **Equation (S7)**, and $\mathbf{R_x}(\alpha)$, $\mathbf{R_y}(\beta)$ and $\mathbf{R_z}(\gamma)$ are the rotation matrices around the [100], [010] and [001] axes by angles $\alpha$, $\beta$ and $\gamma$, respectively. The slices of the simulated 3D intensity distribution for the standard deviations of $\delta\varphi = 10°$, $15°$ and $20°$ are shown in **Figure S10c–e**. As one can see from the figure, the resemblance of the experimentally measured data is much better for the angularly disordered nanooctahedra. Unfortunately, the fact that the experimentally measured form factor is affected by many different factors does not allow to determine the exact values of the standard deviation.

However, we believe the standard deviation is about $\delta\varphi \approx 15°$ as it results in the form factor most similar to the experimentally measured scattered intensity.

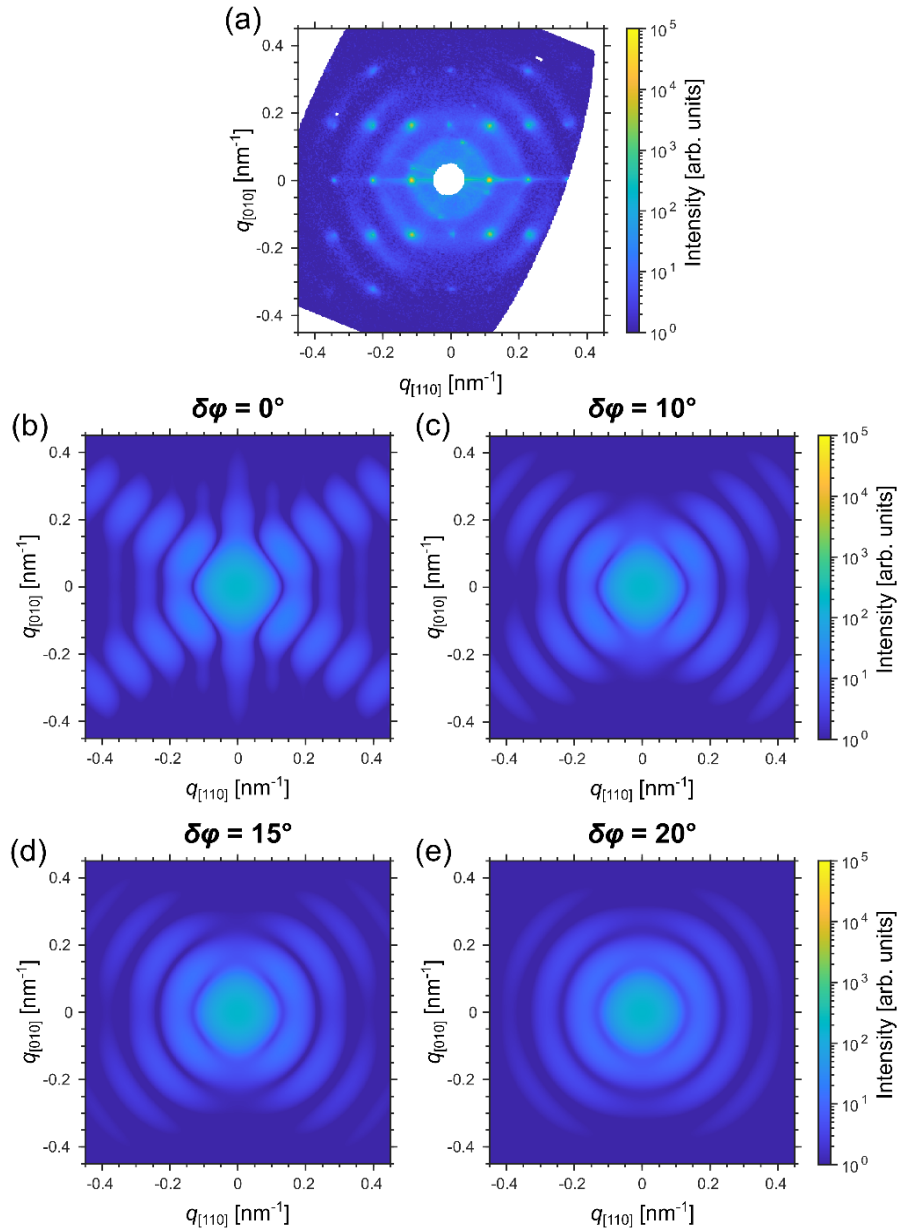

**Figure S10.** (a) Slice through the measured 3D intensity distributions for the O1 sample through the reciprocal space origin and normal to the $[1\bar{1}0]$ axis. (b – e) Slices through the simulated form factor of octahedra with the orientation corresponding to the sketch in **Figure 5** of the main text with the standard deviation of the angular distribution $\delta\varphi = 0°$ (b), $\delta\varphi = 10°$ (c), $\delta\varphi = 15°$ (d) and $\delta\varphi = 20°$ (e). The slices are taken through the reciprocal space origin and normal to the $[1\bar{1}0]$ axis.

Additional slices of reciprocal space normal the [001] and [1$\bar{1}$0] axes for both the experimentally measured scattered intensity and the simulated form factor with the angular standard deviation of $\delta\varphi = 15°$ are shown in **Figure S11** for comparison. We did not simulate the effect of the tip truncation on the form factor because it is out of the scope of this paper.

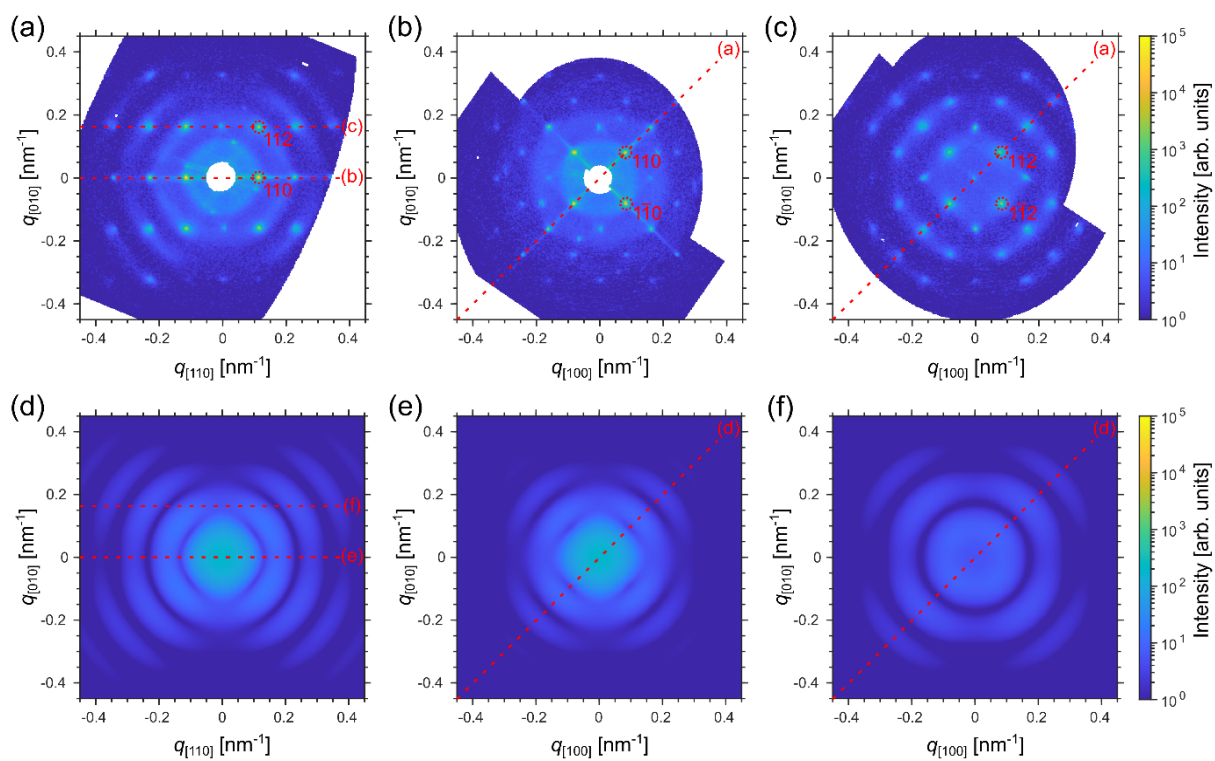

**Figure S11.** (a – c) Slices through the measured 3D intensity distributions for the O1 sample: (a) through the reciprocal space origin and normal to the [1$\bar{1}$0] axis, (b) through the reciprocal space origin and normal to the [001] axis, and (c) through the 002 Bragg peak and normal to the [001] axis (the offset along the [001] axis is 0.162 nm$^{-1}$). Along with the Bragg peaks, anisotropic features of the form factor of the gold octahedral nanoparticles are well visible. (d – f) Slices through the simulated form factor of octahedra with the orientation corresponding to the sketch in **Figure 5** of the main text with the standard deviation of the angular distribution $\delta\varphi = 15°$. The slices are taken (d) through the reciprocal space origin and normal to the [1$\bar{1}$0] axis, (e) through the reciprocal space origin and normal to the [001] axis and (f) through the 002 Bragg peak and normal to the [001] axis (the offset along the [001] axis is 0.162 nm$^{-1}$). The red dashed lines indicate the slices shown in other panels.

Analogous slices of reciprocal space for samples O2 – O5 are shown in **Figures S12.1 – S12.4**. As can be seen from the figure, they are similar to the ones discussed above, but scaled up due to the smaller size of the nanooctahedra constituting these samples. This indicates the same orientation of the nanooctahedra inside the superlattice in these samples.

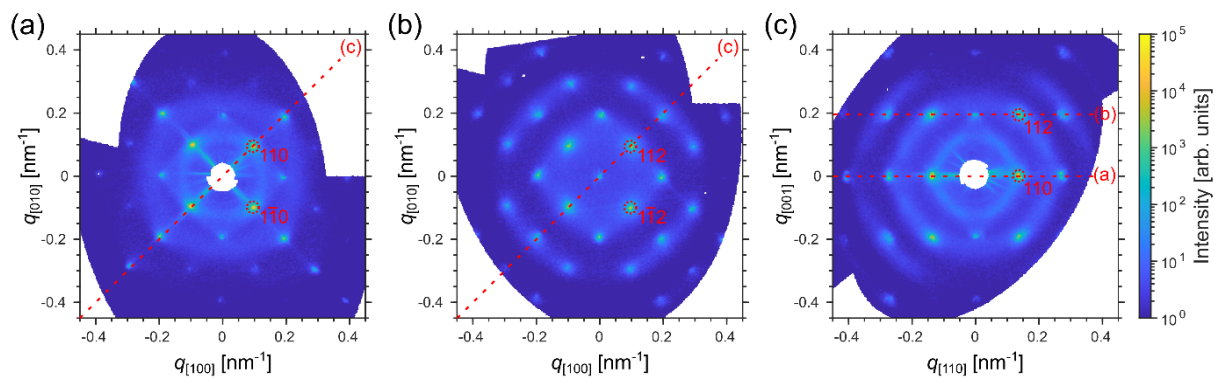

**Figure S12.1.** Slices through the measured 3D intensity distributions for the O2 sample: (a) through the reciprocal space origin and normal to the [001] axis, (b) through the 002 Bragg peak and normal to the [001] axis (the offset along the [001] axis is 0.195 nm$^{-1}$), and (c) through the reciprocal space origin and normal to the [1$\bar{1}$0] axis. The red dashed lines indicate the slices shown in other panels. Along with the Bragg peaks, anisotropic features of the form factor of the gold octahedral nanoparticles are well visible.

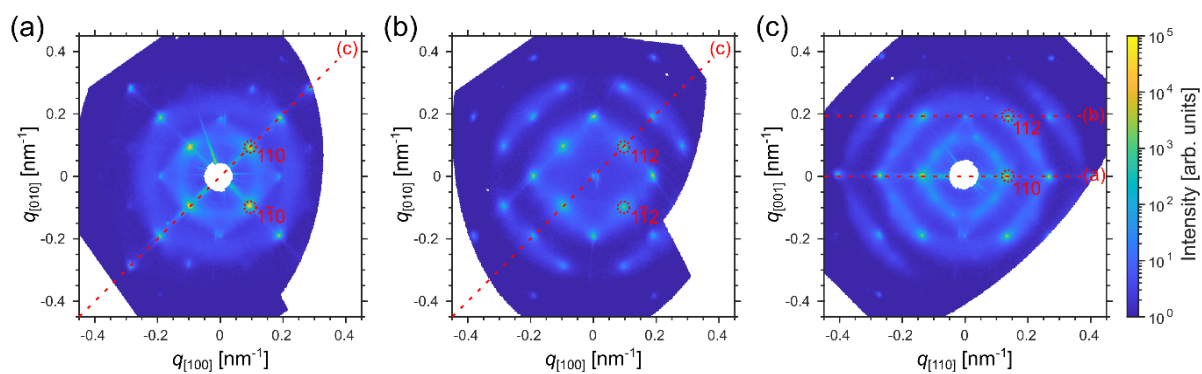

**Figure S12.2.** Slices through the measured 3D intensity distributions for the O3 sample: (a) through the reciprocal space origin and normal to the [001] axis, (b) through the 002 Bragg peak and normal to the [001] axis (the offset along the [001] axis is 0.190 nm$^{-1}$), and (c) through the reciprocal space origin and normal to the [1$\bar{1}$0] axis. The red dashed lines indicate the slices shown in other panels. Along with the Bragg peaks, anisotropic features of the form factor of the gold octahedral nanoparticles are well visible.

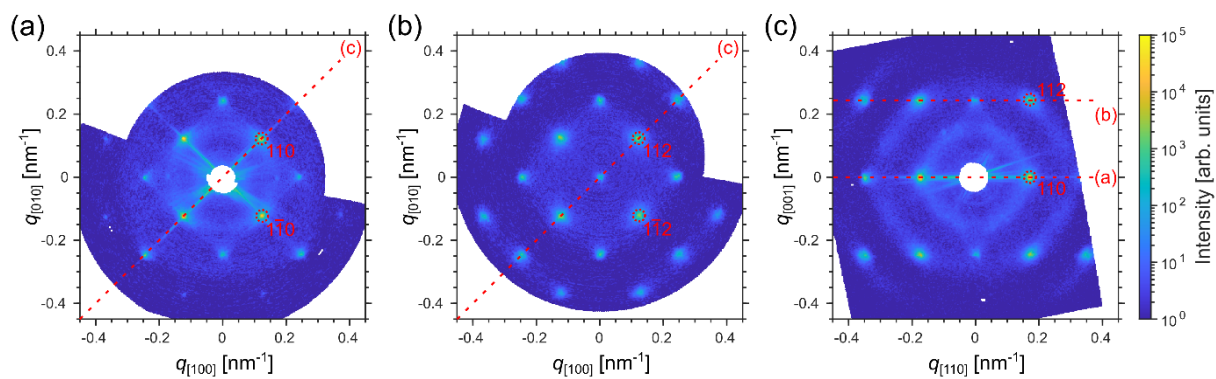

**Figure S12.3.** Slices through the measured 3D intensity distributions for the O4 sample: (a) through the reciprocal space origin and normal to the [001] axis, (b) through the 002 Bragg peak and normal to the [001] axis (the offset along the [001] axis is 0.245 nm$^{-1}$), and (c) through the reciprocal space origin and normal to the [1$\bar{1}$0] axis. The red dashed lines indicate the slices shown in other panels. Along with the Bragg peaks, anisotropic features of the form factor of the gold octahedral nanoparticles are well visible.

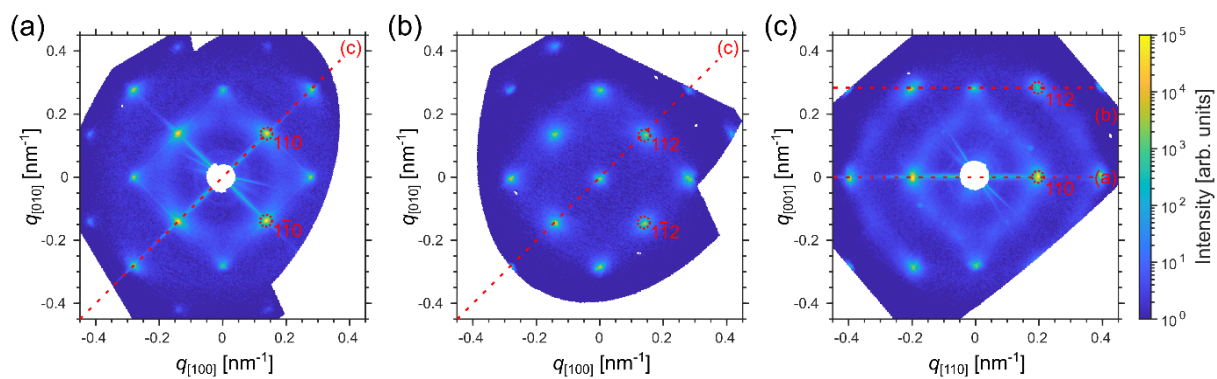

**Figure S12.4.** Slices through the measured 3D intensity distributions for the O5 sample: (a) through the reciprocal space origin and normal to the [001] axis, (b) through the 002 Bragg peak and normal to the [001] axis (the offset along the [001] axis is 0.280 nm$^{-1}$), and (c) through the reciprocal space origin and normal to the [1$\bar{1}$0] axis. The red dashed lines indicate the slices shown in other panels. Along with the Bragg peaks, anisotropic features of the form factor of the gold octahedral nanoparticles are well visible.